\documentclass[sigconf]{acmart}

\AtBeginDocument{\providecommand\BibTeX{{\normalfont B\kern-0.5em{\scshape i\kern-0.25em b}\kern-0.8em\TeX}}}

\copyrightyear{2024}
\acmYear{2024}
\setcopyright{rightsretained}
\acmConference[UIST '24]{The 37th Annual ACM Symposium on User Interface Software and Technology}{October 13--16, 2024}{Pittsburgh, PA, USA}
\acmBooktitle{The 37th Annual ACM Symposium on User Interface Software and Technology (UIST '24), October 13--16, 2024, Pittsburgh, PA, USA}
\acmDOI{10.1145/3654777.3676421}
\acmISBN{979-8-4007-0628-8/24/10}

\usepackage{siunitx}
\usepackage{fdsymbol}
\newcommand{\added}[1]{\textcolor{black}{#1}}
\newcommand{\removed}[1]{}

\newcommand{\system}{\removed{Cardinality}\added{CARDinality}}

\newcommand{\figurewidth}{\linewidth}

\begin{document}

\title{\removed{Cardinality} \added{CARDinality}: Interactive Card-shaped Robots with Locomotion and Haptics using Vibration}

\author{Aditya Retnanto}
\authornote{The first three authors contributed equally to this research.}
\affiliation{\institution{University of Chicago}
  \city{Chicago}
\state{Illinois}
  \country{USA}
}
\email{aretnanto@uchicago.edu}

\author{Emilie Faracci}
\authornotemark[1]
\affiliation{\institution{University of Chicago}
  \city{Chicago}
\state{Illinois}
  \country{USA}
}
\email{efaracci@uchicago.edu}

\author{Anup Sathya}
\authornotemark[1]
\affiliation{\institution{University of Chicago}
  \city{Chicago}
\state{Illinois}
  \country{USA}
}
\email{anups@uchicago.edu}

\author{Yukai Hung}
\affiliation{\institution{National Taiwan University}
  \country{Taipei}
  \country{Taiwan}
}
\email{b09902040@csie.ntu.edu.tw}

\author{Ken Nakagaki}
\affiliation{\institution{University of Chicago}
  \country{Chicago}
\state{Illinois}
  \country{USA}
}
\email{knakagaki@uchicago.edu}

\renewcommand{\shortauthors}{Retnanto, Faracci and Sathya, et al.}

\begin{abstract}
   This paper introduces a novel approach to interactive robots by leveraging the form-factor of cards to create thin robots equipped with vibrational capabilities for locomotion and haptic feedback. The system is composed of flat-shaped robots with on-device sensing and wireless control, which offer lightweight portability and scalability. This research introduces a hardware prototype \removed{and a training approach utilizing a computer vision-based method to explore the possibility of vibration-based omnidirectional locomotion}\added{to explore the possibility of ‘vibration-based omni-directional sliding locomotion'}. Applications include augmented card playing, educational tools, and assistive technology, which showcase\removed{Cardinality's} \added{CARDinality's} versatility in tangible interaction. 
\end{abstract} 

\ccsdesc[500]{Computer systems organization~Embedded systems}
\ccsdesc[300]{Computer systems organization~Redundancy}
\ccsdesc{Computer systems organization~Robotics}
\ccsdesc[100]{Networks~Network reliability}

\keywords{interaction design, flat, robot learning, actuated tangible interface, cards, card robot}

\begin{teaserfigure}
  \includegraphics[width=\textwidth]{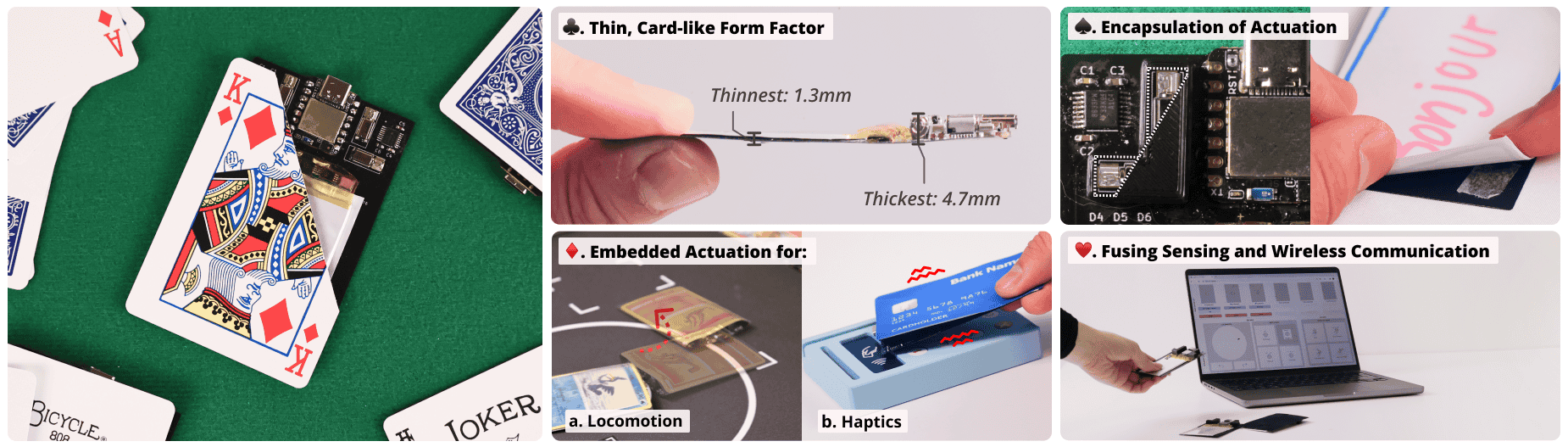}
  \caption{\removed{\textit{Cardinality}} \added{\textit{CARDinality}} features (1) Thin, card-like form factor, (2) Embedded actuation for locomotion and haptics (3) Encapsulation of actuation, and (4) Fusing sensing and wireless communication. }
  \Description{Description of teaser figure}
  \label{fig:teaser}
\end{teaserfigure}


\maketitle

\section{Introduction}

Originating in China circa AD1000~\cite{wilkinson_chinese_1895}, playing cards have evolved into a ubiquitous element across diverse \removed{cultural landscapes}\added{cultures}, offering a plethora of material functionalities including shuffling, stacking, dealing, cutting, fanning, folding and flipping~\cite{altice_playing_2014}. These functionalities are underpinned by intrinsic physical attributes \added{of the card form-factor} such as planarity, uniformity, spatiality and textural properties~\cite{altice_playing_2014}. Such attributes, \removed{in conjunction}\added{along} with their physical characteristics, facilitate a broad spectrum of applications and utility, spanning from recreational pastimes like\removed{playing cards} play to essential tools such as credit cards, flash cards, business cards,\removed{trading cards} and key cards\removed{etc}. As articulated in "The Playing Card: An Illustrated History"~\cite{hoffmann1973playing}, the fundamental simplicity of cards is \added{that} \removed{encapsulated in the statement,}\textit{"Any thin, stiff piece of material can be used as a playing card,"} underscoring their pervasive and uncomplicated nature. 

This materiality extends beyond conventional applications to specialized domains such as structured brainstorming~\cite{roy_card-based_2019}, design methodologies~\cite{hsieh_what_2023}, and educational contexts~\cite{smith2009educational, odenweller1998educational}. Human-Computer Interaction (HCI) researchers have attempted to capture and utilize this ubiquity in contexts such as design toolkits~\cite{hsieh_what_2023}, Augmented Reality (AR) card games~\cite{albert2006art} and educational toolkits~\cite{shen2021value}.
Many of these endeavors involve augmenting traditional physical cards through the integration of digital layers such as AR or incorporating additional sensing technologies directly into the cards~\cite{kirshenbaum_pepa_2018} or the playing surface~\cite{villar2018project}. 

In this document, we explore the incorporation of actuation into the card form factor, resulting in the development of a novel robotic platform. Our objective is to unlock new affordances, applications and interactions by broadening the design space of traditional cards through the introduction of three supplementary layers -- \textit{locomotion}, \textit{sensing} and \textit{haptics}. Although previous research has explored sensing to different degrees, to the best of our knowledge, our distinctive contribution lies in the amalgamation of all three components by utilizing a novel \removed{vibration-based locomotion}\added{\textbf{vibration-based omni-directional sliding locomotion}} approach in a card-like form factor.

To help drive the exploration and development of our cluster of card-shaped robots, we have crafted a set of design and engineering criteria rooted in the inherent capabilities of cards: 

{\begin{itemize}
    \setlength{\itemindent}{-0.7em}
    \item[ $\clubsuit$] \textbf{Thin, Card-like Form Factor:} The device must emulate the slim profile of traditional cards to seamlessly integrate with card-based interactions.
    \item[{\color{red}{$\vardiamondsuit$}}] \textbf{Embedded Actuation for Locomotion, and Haptics:} \removed{Our focus is on incorporating}\added{We incorporate} two primary functionalities: on-table locomotion, and in-hand haptic feedback. These functionalities are aligned with the prevalent usage of cards\removed{, both on surfaces and in-hand}.
    \item[$\spadesuit$] \textbf{Encapsulation of Actuation to Facilitate Customization:} We aim to create a versatile hardware platform conducive to customization. As such, the actuation mechanism must be encapsulated to enable easy integration with existing cards or sleeves, avoiding exposed components such as wheels that could impede customization efforts.
    \item[{\color{red}{$\varheartsuit$}}] \textbf{Fusing Sensing and Wireless Communication:} \removed{To foster unencumbered interaction modalities, t}\added{T}he device will integrate wireless communication and sensing. This holistic approach ensures seamless interaction and communication between the device and its environment.
\end{itemize}}

To tackle these design criteria,\removed{we draw inspiration from Bristlebots\footnote{\url{https://en.wikipedia.org/wiki/Bristlebot}}} \removed{and Kilobots~\cite{rubenstein_kilobot_2012}, leveraging}vibration \added{was chosen over other actuation modalities} as \removed{a}\added{it serves as a} dual-purpose mechanism for both locomotion and haptic feedback in a manner conducive to an encapsulated \added{and thin} design. Following a series of iterative prototypes, we devise a proof-of-concept hardware implementation featuring vibration motors, a Bluetooth Low Energy (BLE)-based microcontroller equipped with an Inertial Motion Unit (IMU), a battery, and other essential components. These components are integrated onto a semi-flexible rectangular Printed Circuit Board (PCB), ensuring compatibility with card-based interactions.  

In the rest of this document, following a literature review (Section~\ref{section:related-work}), we introduce a design space (Section~\ref{section:design-space}) laying out the Input/Output (I/O) capabilities of the \system{} platform. 
The design space is intended to serve as a library that can be employed when designing various versatile applications. 

Subsequently, the implementation section (Section~\ref{section:implementation}) delves into the details of our proof-of-concept, encompassing both hardware and software aspects aimed at controlling and programming the diverse functionalities of our device. Our locomotion system controls the operation of multiple vibration motors in varied configurations to facilitate omni-directional movement. To accomplish this, we explore and refine how different vibration configurations influence locomotion, employing a computer vision-based closed-loop training setup (Section~\ref{section:training}). Following the training phase, the device is \removed{endowed with the capability to execute}\added{capable of} omni-directional locomotion\removed{, traversing arbitrary directions or paths}. Apart from encapsulating the actuation, using vibration-based locomotion allows us to explore omni-directional motion\removed{ as} compared to a regular differential driven DC motor design. 

In Section~\ref{section:tech-eval}, we undertake an evaluation of the robustness and transferability of our training process\removed{,}\added{. Section ~\ref{section:applications}} showcas\removed{ing}\added{es} a wide range of applications, including card games, educational tools, and other card-based \removed{everyday }activities. Finally, we close our discourse with a comprehensive discussion (Section~\ref{section:discussion}), shedding light on the limitations of our work and\removed{delineating} potential\removed{avenues for} future research in this domain.

Our contributions include:

\begin{itemize}
    \item A general approach to build card-shaped robots with vibration-based actuators, serving both on-table \textit{locomotion,} and in-hand \textit{haptics} with integrated sensing. 
    \item Proof-of-concept hardware implementation \added{with supplementary training set up and software methods for \textbf{vibration-based omni-directional sliding locomotion}}.
    \removed{\item Training setup and software methods for vibration-based locomotion along with a technical evaluation of our approach.} 
    \item A range of applications that demonstrate the unique interaction capabilities of \system{}.
\end{itemize}
 \section{Related Work}
\label{section:related-work}

\begin{figure*}[t]
    \centering
\includegraphics[width=\figurewidth]{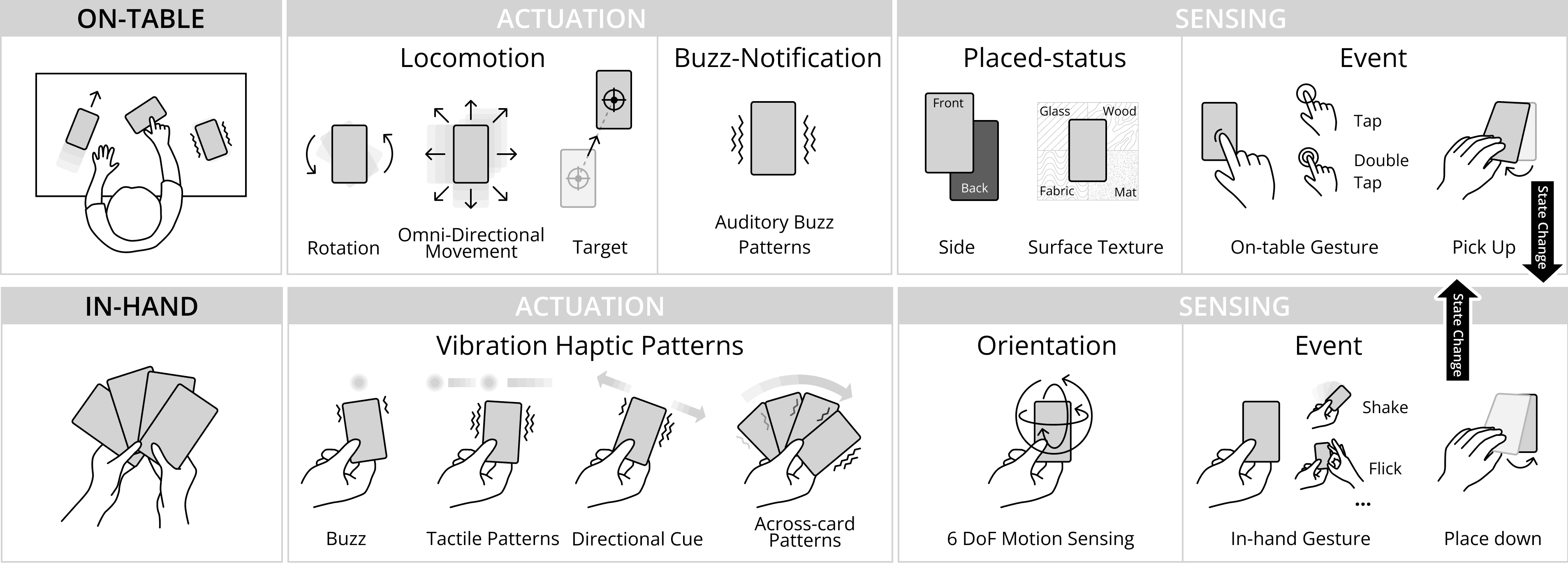}
    \caption{Design Space of \removed{Cardinality}\added{CARDinality}, divided across On-Table and In-Hand status, where each status has diverse Actuation and Sensing capability based on the affordance of cards.}
    \label{fig:designspace}
\end{figure*}

We outline prior works in (1) cards in HCI, (2) actuated TUIs and robots in various thin forms, and (3) vibration-based locomotion. 

\subsection{Cards in HCI}

Cards are \removed{ubiquitously present }\added{widely adopted} across cultures, resulting in \removed{a heavy presence}\added{use cases such as}\removed{ in} modern tabletop gaming,\removed{along with} educational uses \cite{scalise_deal_2022, knapp_learning_1996, golick_deal_1973} \added{and fortune-telling~\cite{place_tarot_2005}}.\removed{Cartomancy~\cite{place_tarot_2005} -- the use of cards for fortune-telling and divination-- is also heavily prevalent in cultures across the world.} Their prevalence is tied to the simplicity of their design and the versatility of their uses. Over time, the versatility of cards has invited the embedding of technology and computing within them. Poker games broadcasted over television or the internet make use of smart cards with RFID tags~\cite{romer_smart_2002} to communicate the rank and suit of the card through a smart table to the live audience. In a similar vein, \removed{multiple }researchers have explored systems and use cases for smart cards~\cite{rankl_smart_2004, schnorr_efficient_1991}, especially in\removed{ use} cases related to security~\cite{mayes_smart_2008}.

Within HCI research, cards have a strong prominence as design tools~\cite{hsieh_what_2023} as they are ``simple, tangible and easy to manipulate''~\cite{roy_card-based_2019}. These methods are used by practitioners in\removed{ the} industry~\cite{noauthor_method_nodate} and academic researchers alike~\cite{friedman_envisioning_2012, rogerson_smeft_2022, lomas_design_2021, deng_tango_2014, lucero_plex_2010, elsayed-ali_responsible_2023, fedosov_sharing_2019, alves_deck_2011, caraban_nudge_2020}. Researchers have also proposed prototypes of cards that bring interactivity to play decks. Kirshenbaum et al. \removed{speculate and }propose an early prototype -- PEPA (Paper-like Entertainment Platform Agents)~\cite{kirshenbaum_pepa_2018}. Along similar lines, FluxPaper~\cite{ogata_fluxpaper_2015} \removed{proposes}\added{presents} the addition of a patterned magnetic layer on paper to enable physical movement using a magnetic field. Researchers also\removed{propose} \added{suggest} interaction techniques~\cite{holman_paper_2005} for digital paper, some of which can be transferred to card-based interfaces. Apart from\removed{proposing} potential designs and uses for cards, researchers have also \removed{proposed the use of}\added{pointed to} external novel actuators -- such as shape displays~\cite{ishii_transform_2015} or tabletop robots \cite{toioblackjack} -- to\removed{ handle and} move cards on\removed{ a tabletop} surface\added{s}. In contrast, we explore the design and engineering of a fully embedded system where locomotion, sensing, haptics, and wireless communication are all \removed{embedded}\added{fused} into a single \removed{device in a }thin, card-like form factor.

\subsection{Robotic TUIs with Various (Thin) Material Form-Factors and Thin Robots}

HCI researchers have been interested in developing robots and robotic interfaces that often derive inspiration from common material form factors in the real world. While LineFORM~\cite{nakagaki_lineform_2015}\removed{derives} \added{takes} inspiration from lines and curves to create an Actuated Tangible UI that physicalizes digital curves, ChainFORM~\cite{nakagaki_chainform_2016} derives inspiration from tapes. Fiber-material-inspired actuated interfaces have \removed{also }been developed \cite{kilic_afsar_omnifiber_2021,  kolvenbag_rapid_2022} to fuse the affordance of weavable fiber-like materials with robotic actuation for haptic and tangible interactions. Swarm robots such as Zooids~\cite{le_goc_zooids_2016} are also available in small form factors to provide cluster-based affordances. \removed{In another tangent, r}\added{To actuate thin, paper, and paper-like material, r}esearchers have \removed{also }proposed the inclusion of Shape Memory Alloys (SMAs) and other heat-reactive materials\removed{to actuate thin, paper, and paper-like material}~\cite{koizumi_animated_2010, roudaut2013morphees, heibeck2015unimorph, gomes2013morephone}, adding\removed{movement and} animation to thin substrates. 

In the robotics domain, thin, flat locomotive robots have been explored for building autonomous systems that navigate and explore environments with narrow paths. Such hardware employs different actuation methods such as inflatables ~\cite{shepherd2011multigait}, or foldable origami ~\cite{belke2017mori, firouzeh2015robogami, felton2014method}. 

\removed{While some of these works in both HCI and robotics utilize the available flat form factors mostly as a transitional state,} \added{Our work} \removed{we specifically}focus\added{es} on the flatness in order to develop an interactive robotic system that uses the form factor of cards \added{whereas prior work uses flatness as a transitional state}. We utilize thin, semi-flexible PCBs and focus our ideation and prototyping around adding locomotion and haptics to cards in an encapsulated form factor. 

\subsection{Vibration-based Locomotion\removed{and Bristle Bots}}

While traditional robots naturally utilize deterministic methods of locomotion to maintain precision, some robots utilize stochastic methods\removed{in specific use cases}. Bristle bots are popular toys that utilize vibration to create fun\removed{and}\added{, }fast \added{and random} locomotion for children to engage with. \removed{While the locomotion is fast due to the vibration being transferred to the bristles at the bottom, it is entirely random and not controlled in any way.} Researchers have attempted to control this form of locomotion. Kilobot~\cite{rubenstein_kilobot_2012} utilizes two sealed coin-shaped vibration motors commonly used by haptics researchers to enable locomotion in a swarm setting. Ratchair~\cite{parshakova_ratchair_2016} also utilizes vibration as a mechanism at a much larger scale to move a piece of furniture from one location to a predetermined destination purely utilizing vibration. Other devices also rely on thinner actuators like piezo-electric actuators~\cite{cheng_eviper_2023} even on a millimeter scale~\cite{hao_controlling_2023}. Such microbristle bots have been proposed\removed{ as a method} to perform\removed{ a multitude of} tasks like pipe inspection and microsurgery. 

The \removed{natural}advantage of a vibration-based system is that the actuator does not need to directly make contact with a surface\removed{in order} to translate the applied force into motion. Vibration\removed{essentially} lifts the device on a micro-scale and pivots it on the opposite corner to move the device in a particular way.\removed{Apart from allowing us to provide haptic sensations when the card is picked up,} \added{Vibrations can provide haptic sensations and}\removed{this enables us to} fully encapsulate the card\removed{,}\added{.} \added{This} let\removed{ting}\added{s} users grasp it at any location without any interference from an actuator \added{and enable customizations}. 

In legged approaches such as kilobots, \added{and} Ratchair \removed{, and Bristle bots}vibration locomotes the robots through a stick-slip manner. The legs bias the robots to certain directions and forces the robot to \removed{locomote to}move inch by inch. The affordance of cards discourages legs and thus our robot is \textit{legless}. When designing \textit{how} a card-shaped should locomote \textit{sliding}\removed{and \textit{gliding}} is the natural gait. Combining the legless design with vibrations allows omni-directional motion compared to the differential movement commonly utilized by wheel-driven robots and other vibration-driven robots. In order to control this stochastic locomotion method, we develop a training system that lets the robot `learn' how to move in an omni-directional manner based on different vibration configurations. This approach was preliminarily explored in Ratchair to make chairs locomote based on two large vibration motors~\cite{parshakova_ratchair_2016}, but we extend this approach to apply for thin, mobile robots, targeting to achieve omni-directional movement. \section{CARDinality Design Space}
\label{section:design-space}

\begin{figure}[h]
    \centering
\includegraphics[width=\figurewidth]{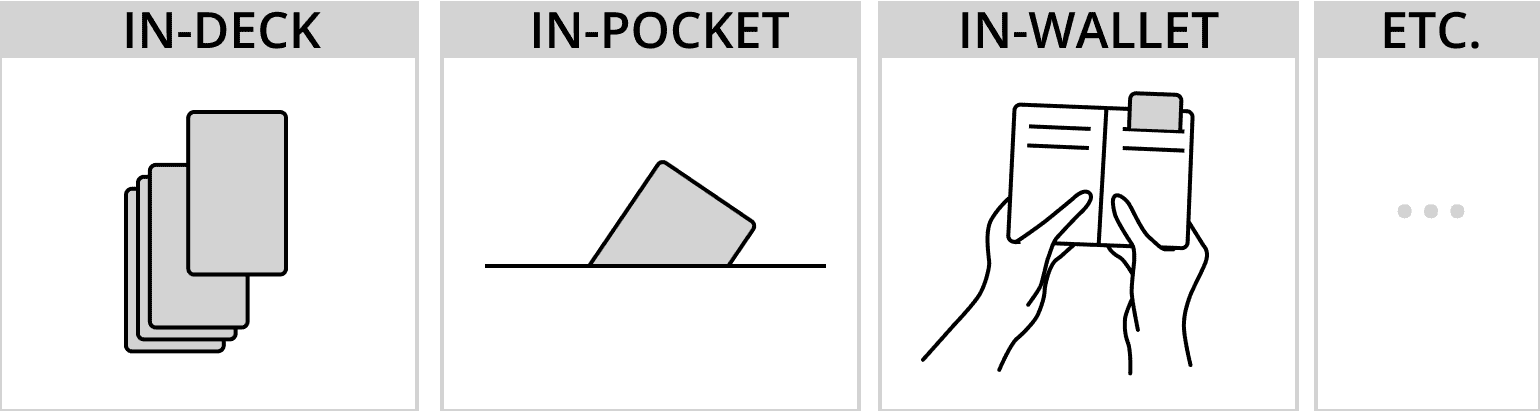}
    \caption{Other states for the CARDinality device}
    \label{fig:other states}
\end{figure}

In this section, we outline the overall capabilities of \system{} with a design space, illustrated in Figure \ref{fig:designspace}. 
Card-shaped objects can be generally conceived\removed{of} as being in two major states: \textbf{1. On-Table} (Section~\ref{section:on-table}) - when cards are placed on tabletop surfaces, and \textbf{2. In-Hand} (Section~\ref{section:in-hand}) - when cards are being held in the user's hands. 
Both these primary states serve explicit usage modalities and affordances. \added{Throughout our research process, we also preliminarily explored other states, including \textbf{In-Pocket}, \textbf{In-Wallet}, and \textbf{In-Deck} (Figure~\ref{fig:other states}). These states provide a suite of additional interactions and novel research prospects, some of which are explored in Section~\ref{section:applications}.}

\removed{For example, in many games, cards placed on the table are usually in a discard pile or a draw pile. The draw pile refers to cards that can be picked up by players during their turn and the discard pile refers to the cards that have already been discarded by the players.} Cards placed\removed{on the table} \added{On-Table} can\removed{also} be positioned face up, which invites all participants to strategically use these cards during the game\added{, or face down, waiting to be picked up by a player}. 
Meanwhile, cards held\removed{in the player's hand} \added{In-Hand} are often private\removed{and are meant for the player's eyes only}.
Within these two major states, \system{} offers actuation and sensing to create interaction opportunities with users.
\removed{These are detailed in further subsections.}
All actuation capabilities are handled by the vibration motors and most of the sensing capabilities are handled by the IMU. 
The surface detection is an exception that uses \added{both} the vibration motors and IMU in conjunction with each other. 

\removed{While there can exist other states for cards, such as \textit{stacking}, or \textit{fanning}, in this paper, we focus on these two major states as an initial exploration as further states can be built upon these base states based on the application being considered.} 

\subsection{On-Table}
\label{section:on-table}

\textbf{Actuation:} \removed{While i}\added{I}n the On-Table state, \system{} leverages the vibration motors to afford omni-directional \textbf{Locomotion}. In our work, we\removed{solely} focus on \textit{Omni-Directional} movement and \textit{Rotation}. A target-based closed-loop locomotion system is\removed{ also} possible by adding an external camera. Our robotic platform is capable of individual movement \removed{as well as}\added{and} facilitating swarm-like clustered interactions. The vibration motors also serve a second purpose emitting \textbf{\removed{Buzz-Signals}\added{Buzz-Notifications}}, effectively notifying users and capturing their attention through the auditory function of vibration.

\textbf{Sensing:} Using the on-board IMU, our system is designed to discern certain sub-states and events while in the On-\removed{t}\added{T}able state. Firstly, it can reliably detect the orientation of the card, differentiating between its \textbf{Side} being face-up or face-down on the table surface. Secondly, in conjunction with the vibration motors, we explore the development of a classifier aimed at identifying the \textbf{Surface Texture} upon which the card is placed. Additionally, our system \removed{is capable of detecting}\added{can detect} other on-table events, such as tap gestures, thereby expanding its range of interactive capabilities. 

When the card is picked up, it transitions to the In-Hand state described in the next section.

\subsection{In-Hand}
\label{section:in-hand}

\textbf{Actuation:} When the robot is held in the user's hand, \system{} leverages the vibration motors to generate haptic patterns, facilitating the transmission of information and providing feedback to the user. Analogous to how specific information about cards is confined to the individual holding the card, \textbf{Haptics} are conveyed using the vibration motors, thereby enabling the delivery of private information exclusively to the person holding the card. 

\textbf{Sensing:} The on-board IMU can be used to offer 6 Degrees of Freedom\removed{ (DoF)} input by accurately sensing the \textbf{Orientation} of the device, thus enabling state switching. Additionally, while in this state, \system{} is capable of detecting various in-hand \textbf{Gestures} such as \textit{shaking} or \textit{flicking} the card, further enhancing its interactivity.

 \section{Implementation} 
\label{section:implementation}
\begin{figure}[h]
    \centering
\includegraphics[width=\figurewidth]{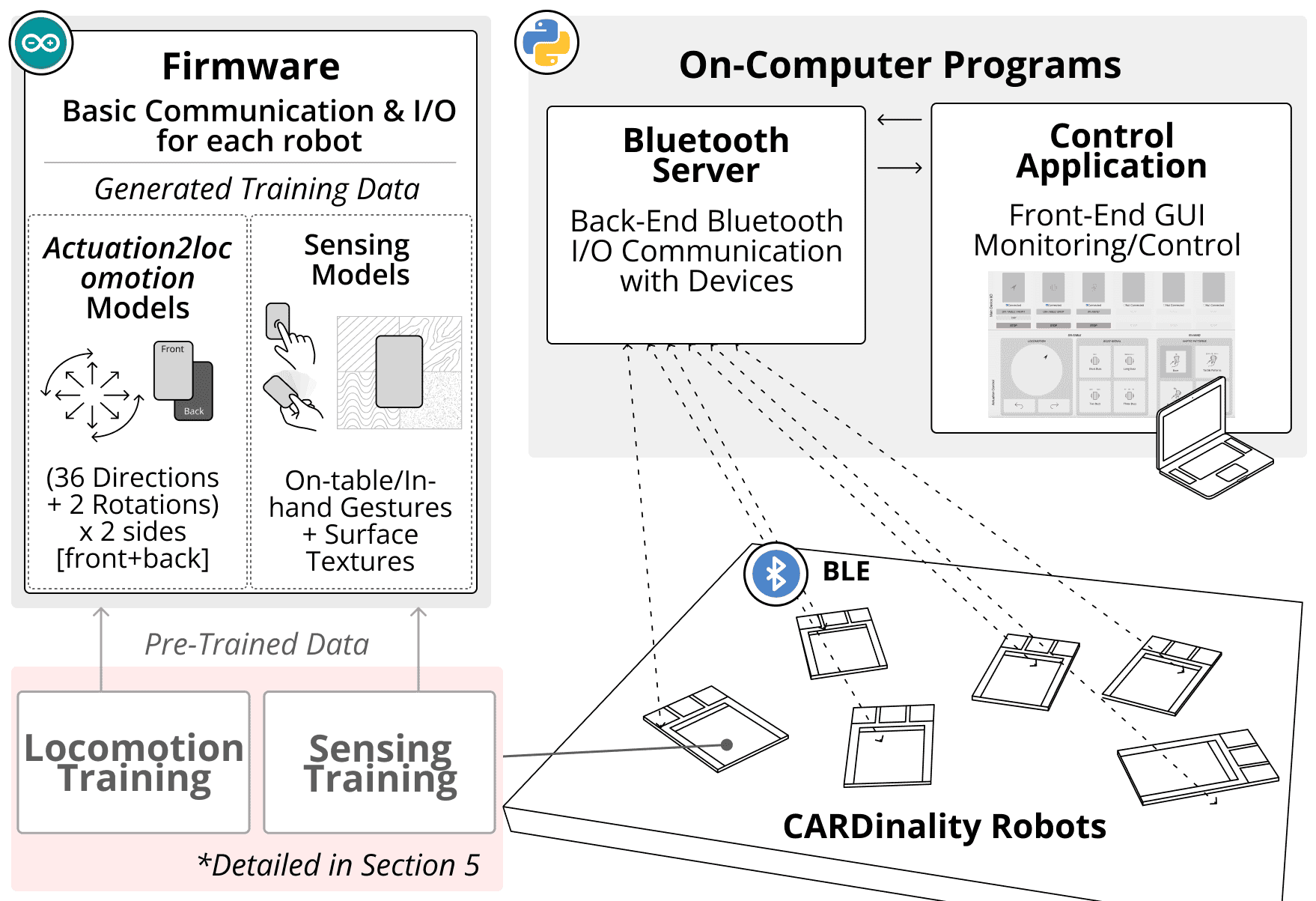}
    \caption{Overall System}
    \label{fig:overallsystem}
\end{figure}

In this section, we outline how our proof-of-concept system is built. Our overall system comprises the \system{} robot's hardware and a software stack that controls and monitors the devices as described in Figure \ref{fig:overallsystem}. The specific parameters and the process to derive these parameters for each robot are described in Section~\ref{section:training}.

\subsection{CARDinality Hardware}

\begin{figure}[h]
    \centering
\includegraphics[width=\figurewidth]{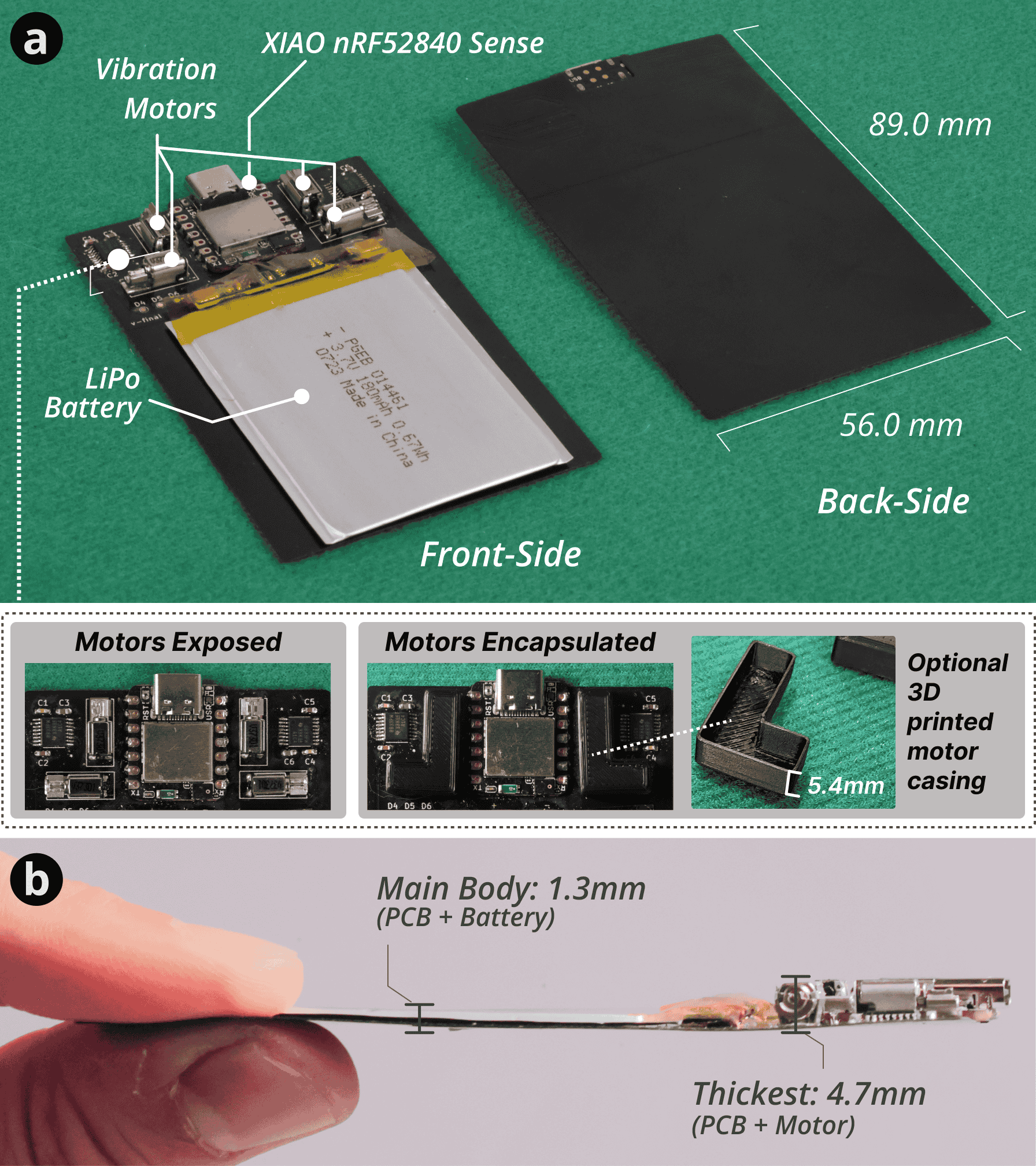}
    \caption{\removed{Cardinality} \added{CARDinality} Robot (a) hardware overview including motor encapsulation states, (b) hardware thickness.}
    \label{fig:hardware}
\end{figure}

Figure \ref{fig:hardware} provides a detailed depiction of the \system{} robot. 
A semi-flexible PCB, measuring $0.3 \unit{\milli\metre}$ in thickness was designed by us and manufactured by PCBWay\footnote{\url{https://www.pcbway.com/}}. 
During initial testing phases, various PCB thicknesses were evaluated, including a standard flexible PCB. 
Results indicated that thicker PCBs ($>0.3\unit{\milli\metre}$) deviated significantly from the desired card form factor, while fully flexible PCBs absorbed vibrations to an extent that impeded the robot's locomotion. 
The dimensions of the PCB are $56 \times 89 \unit{\milli\metre}$, closely resembling those of a standard playing card ($63 \unit{\milli\metre \times 89 \unit{\milli\metre}}$).

The Seeed Studio XIAO NRF52840 Sense microcontroller is employed in the system, offering an array of onboard features such as BLE, IMU, and a LiPo charging circuit, all within a compact and sleek form factor. Notably, the inclusion of the LiPo charging circuit enables streamlined functionality, eliminating the need for multiple connectors on the PCB and facilitating both charging and programming via a single USB-C port. Additionally, the device comprises a slim ($1 \unit{\milli\metre}$) 3.7V LiPo battery with a capacity of 180mAh ($40 \times 60 \times 1 \unit{\milli\metre}$), complemented by a battery protection circuit, 2x dual-channel DC motor drivers (DRV8833C) for motor control, and 4x Eccentric Rotating Mass (ERM) vibration motors. The motor's maximum z-dimension is $4.4 \unit{\milli\metre}$, resulting in the device's thickes point measuring $4.7 \unit{\milli\metre}$, while its thinnest point (Figure~\ref{fig:hardware}b) measures $1.3 \unit{\milli\metre}$. By comparison, a standard playing card typically measures around $0.3\unit{\milli\metre}$ in thickness.

\begin{figure}[h]
    \centering
\includegraphics[width=\figurewidth]{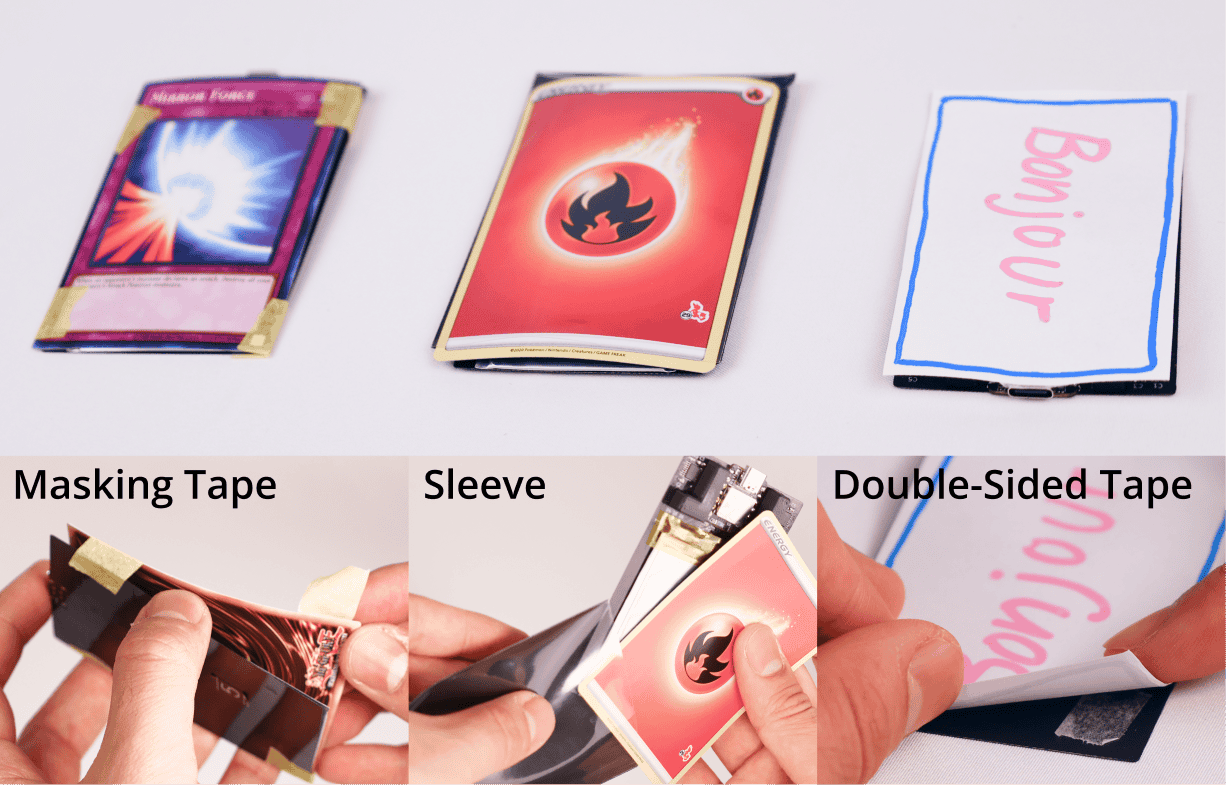}
    \caption{Hardware Customization Examples}
    \label{fig:hardwareCustomization}
\end{figure}

\subsubsection{Motor Selection and Placement}

Our primary objective in the design process was to develop an exceptionally thin device while meeting the requirements outlined in our design space. Since vibration motors and similar actuators are not conventionally utilized for locomotion, we underwent numerous iterations and explored various options during the prototyping phase.

Piezoelectric actuators emerged as a potential solution due to their promise of achieving extreme thinness compared to traditional actuators. 
However, incorporating them posed additional challenges, notably the need for additional legs to direct the actuator's vibration onto the surface to create locomotion.
This deviated from our design criteria.
Similarly, coin-shaped ERM motors, utilized in Kilobots~\cite{rubenstein_kilobot_2012}, appeared promising due to their enclosed form factor.
However, their requirement for perpendicular placement to effectively translate vibration into locomotion increased the device's thickness while offering only modest vibration force.

After careful consideration and testing, we select the Vybronics VZ43FC1B5640007L, a surface-mount ERM vibration motor, for our final design. This choice was based on its compact form factor and its vibration force of $0.65 \unit{\gram}$. 
Smaller ERM motors, while thinner, were deemed inadequate in generating the necessary force for the robots to locomote effectively.

The height of the chosen motor, at $4.3 \pm 0.2 \unit{\milli\metre}$, is slightly lower than that of the microcontroller ($4.8 \unit{\milli\metre}$), serving as a limiting factor in the device's thickness. While it is theoretically possible to reduce the device's height further by employing a microcontroller without a USB-C port, such as the MDBT50Q, our testing revealed that using smaller motors to match this reduced thickness diminished the available vibration force, resulting in weaker locomotion. 

The motor pairs are strategically positioned in an L shape (as depicted in Figure~\ref{fig:hardware}a), allowing the centre of mass to shift towards the selected x-y directions when the motors are activated. The precise placement and orientation of the motors significantly influence locomotion trajectories, and we arrived at the final placement configuration through multiple rounds of trial and error. Furthermore, the placement \removed{takes into consideration}\added{considers} the overall usability of cards, with motors positioned as close to the microcontroller as possible to maintain thin edges on all four corners where cards are traditionally held.

\subsubsection{Customization}

The enclosed design, coupled with the strategic arrangement of thicker components around the microcontroller, facilitates tailored customization of the cards\removed{ as required}. 
Customizations may involve affixing regular or handmade cards onto the robot or inserting the entire robot into commercially available playing card sleeves alongside a regular card.
\removed{The utilization of v}\added{V}ibration-based actuation demonstrates a clear advantage\removed{in this context, as it ensures }\added{, ensuring} that locomotion and haptic feedback remain\removed{s} largely unaffected by\removed{ such} modifications. While these alterations may\removed{ potentially} impact locomotion accuracy, our training pipeline as explained in Section~\ref{section:training} enables us to identify new input control parameters to uphold a consistent level of control\removed{ notwithstanding the modifications}. 

\subsection{Software}
The software architecture comprises three major modules: the robot's firmware, a Bluetooth server, and a Control Application (Figure~\ref{fig:overallsystem}). 

\subsubsection{Firmware for Locomotion and Sensing}

In addition to facilitating Bluetooth communication, the on-device firmware stores and employs the \textit{actuation2locomotion} model and the \textit{sensing model}.
\removed{Typically, firmware is static in nature. 
However, in our novel approach, we adopt a dynamic firmware strategy that adapts different \textit{locomotion} and \textit{sensing} models based on the specific device configuration and any applied customizations.} \added{These models are added in a configuration file, allowing for individualized \textit{locomotion} and \textit{sensing}.} These models are further elaborated in Section~\ref{section:training}.

The \textit{actuation2locomotion} model encompasses 76 motor configurations, including 2 sets of 36 configurations for omni-directional motion in 10-degree increments \removed{(face-up and face-down)} and \removed{2 sets of}rotation configurations for clockwise and counterclockwise movements (face-up and face-down). The \textit{sensing model} (Figure~\ref{fig:training-sensing}) utilizes the onboard IMU (LSM6DS3), to ascertain the device's state (Section~\ref{section:state-detection}), in addition to considering user inputs and environmental factors such as \added{surface detection}\removed{detecting the surface on which the card robot was placed} (Section~\ref{section:surface-detection}).

\subsubsection{Bluetooth Server}

This module is responsible for managing read/write operations to and from the robot. Users can issue either raw motor commands (8-byte instruction) or reprogrammed configurations (1-byte instruction). Additionally, the server handles state, surface, and gesture classifications and raw 3-axis gyroscope data (10-byte messages). During our development, we have verified that our computer (MacBook Air 2021 M1) can establish simultaneous connections with up to 10 \removed{Cardinality} \added{CARDinality} robots without encountering performance issues. 

\subsubsection{Control Application}

\begin{figure}[h]
    \centering
\includegraphics[width=\figurewidth]{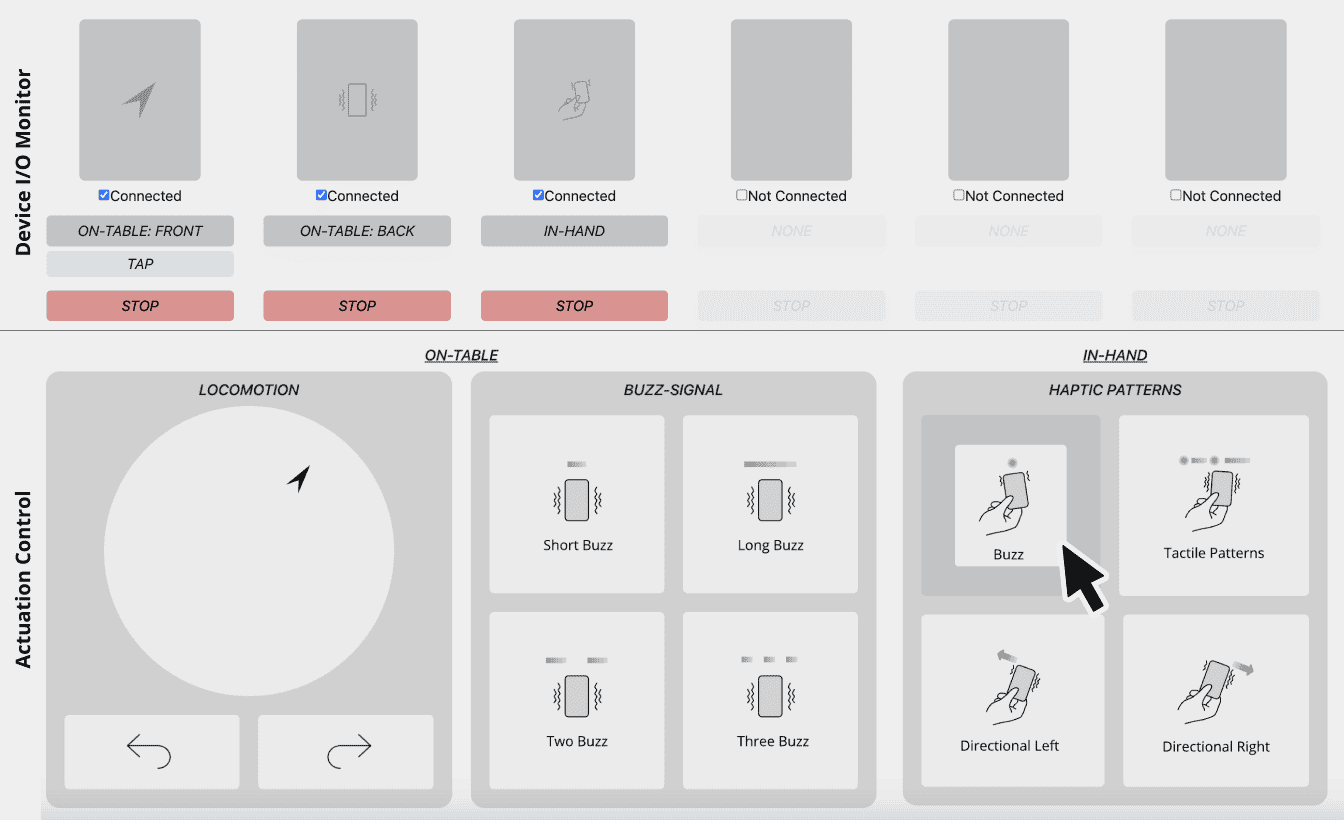}
    \caption{GUI control application for \removed{Cardinality} \added{CARDinality}, with main device I/O and actuation control}
    \label{fig:gui}
\end{figure}

To easily control, monitor, and prototype the behavior of the \system{} robots, we\removed{ also} developed a \added{Javascript-based} GUI application using a Flask server in Python, which communicates to a I/O Handler\removed{ This Flask server is connected to HTML, CSS, and Javascript files to facilitate the GUI}. As shown in Figure \ref{fig:gui}, the top part of the GUI helps users handle BLE connection to \system{} robots, up to 10, and manage I/O for each robot, monitoring the sensing events received from each robot. In this example, three robots are connected to the GUI: one is on-table face-up receiving a tap sensing event, another is on-table face-down, and a third robot is in-hand. The bottom of the GUI is\removed{ specifically} for accessing and configuring a variety of actuations, and users can either use omni-directional input to control the locomotion or select haptic patterns from preset vibration patterns. In this figure, a simple haptic buzz is selected as the current action.

\removed{Additionally, we also constructed a computer vision-based closed-loop control system for the GUI application. When launching this feature, users can click on the camera view window to provide a specific target (x,y) position for the robot (Figure \ref{fig:wirelesscharging}). With the use of Aruco markers, the computer vision tracks the current device position (x,y,deg). This is used to calculate \removed{the omni-directional movement the robot should perform,} \added{a vector between the robot and target to select the discrete omni-directional movement the robot should perform}. \removed{and in turn actuates the device} \added{The robot actuates} using the corresponding pre-trained motor patterns. \added{Targetting is not a major contribution in this work, and will be explored in the future works section.} \removed{In this case, the position of the current device can be actively compared to the target point to aim at the target continuously.}}

\removed{The GUI application provides useful and easy ways to control and test \system{} robots' functionality. Optionally, for users with programming experience, they can also customize motion, and interaction for the robots by coding to create \textit{if X, then Y} via Python.The GUI application should still serve as a useful tool to quickly test out different features before they program the behaviors.} %
 \section{Building the Locomotion and Sensing Models}
\label{section:training}

The training process develops the input and output capabilities of \system.\removed{\textit{Gliding} is the embodied locomotive gait of cards rather than inching and crawling} For our system, we design the robot such that it \removed{moves in a smooth omni-directional manner}\added{is capable of \textbf{vibration-based omni-directional sliding locomotion}}.\added{We seek a mapping that transforms individual motor inputs to 36 discretized omni-directions and 2 rotations (clockwise and counterclockwise.}
\removed{We simplify our approach by discretizing omni-directionality to 36 directions and seek a mapping that transforms individual motor inputs to each direction. Furthermore, we also identify 2 motor patterns that result in a clockwise and counterclockwise rotation. \quad{We said this in implementation}}To the best of our knowledge, there is no analytical model analogous to our approach with\removed{resepect} \added{respect} to planar locomotion.
\removed{Thus} \added{To explore potential for locomotion}, we developed a \added{supplementary} computer vision-based method to empirically explore and optimize input motor patterns for \textit{each side of the robot}. Once found, our mappings $\boldsymbol{\Phi}$ transform 36 motor patterns to an output direction without external peripherals. \removed{Due to the nature of our robot, r}\added{R}otation is easily achieved\removed{ and thus}\added{, so} we only seek motor patterns that\removed{ result in a rotation that } minimize\removed{s} battery usage. In total, we retain 76 motor patterns. For sensing, we outline the specific model and the data pre-processing steps and illustrate them in a state diagram (Figure \ref{fig:training-sensing}). 

\subsection{Planar Locomotion}

We developed an empirical approach to learning \removed{planar locomotion}
\added{vibration-based omni-directional sliding locomotion} for our device by treating the training process like a ``black box''. \system \added{\,} has 4 motors that are parameterized by the motor intensities and direction of rotation of each motor. Motor inputs range from -3.3V to 3.3V with the sign representing the rotation of the motor. We represent this in a length 8 array $\boldsymbol{m}$ \added{ 
representing 4 motors, where each pair of consecutive bytes indicates the counter-clockwise and clockwise intensities for a single motor, respectively.}\removed{ = 
[m1_{cc}, m1_{cw}, m2_{cc},m2_{\removed{c2} \added{cw}}, m3_{cc}, m3_{cw}, m4_{cc}, m4_{cw}].Where cw and cc are clockwise and counterclockwise.} We model locomotion through the \removed{below }function\removed{,} $\boldsymbol{f}$: 
$$
\boldsymbol{P_{n}} = \boldsymbol{f}_{\Phi}(\boldsymbol{m},\boldsymbol{n})
$$
Where $\boldsymbol{n}$ is a time-based sample, and $\boldsymbol{P}$ is an output trajectory consisting of the robot's pose $\{x_{n}, y_{n}, \theta_{n}\}$. 
To balance optimization through $\sim$68 billion combinations of motor intensities, we first utilized a grid search approach to explore the feasibility of \removed{omni-directional locomotion} \added{vibration-based omni-directional sliding locomotion}. In our approach\removed{, although simplistic,} we assume that locomotion is only influenced by motor patterns and that the environment or peripherals have minimal effect on locomotion. Discussion on external factors will be done in the technical evaluation section. 

\subsubsection{Training Set Up}
\begin{figure}[h]
    \centering
\includegraphics[width=\figurewidth]{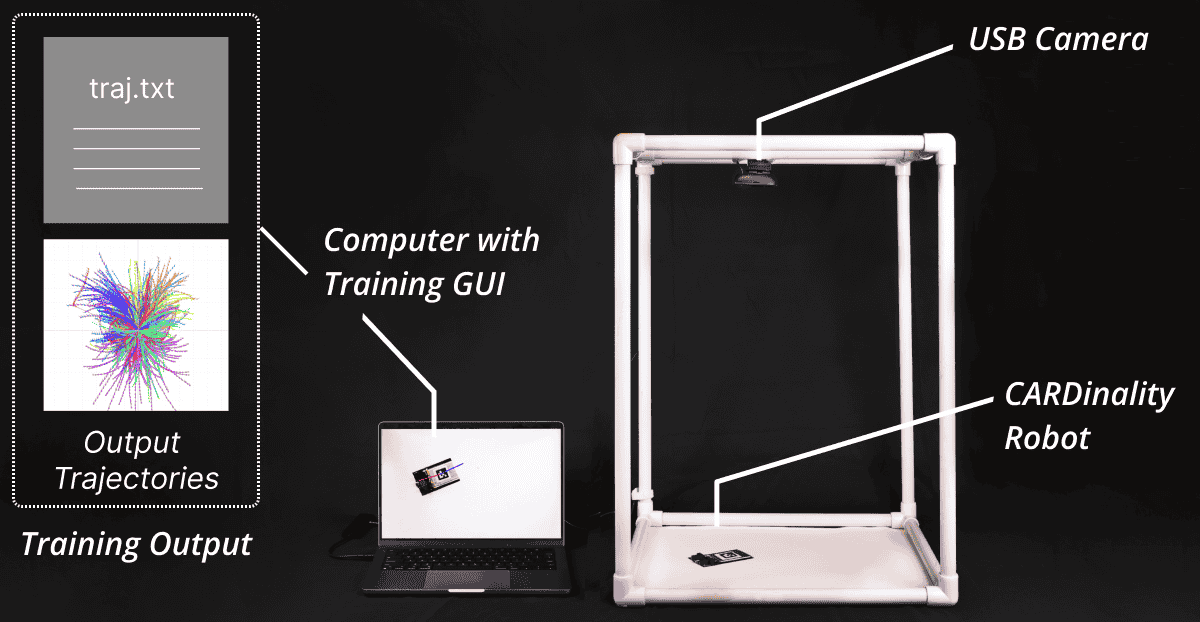}
    \caption{Training setup to find the 36 vectors, which includes a USB camera to track the movement of the \system{} robot}
    \label{fig:trainingsetup}
\end{figure}

We developed a ``robot school'' that encloses the robot into a fixed $44.5 \unit{\centi\metre} \times 36\unit{\centi\metre}$ space. A camera is placed on the top of the enclosure and Aruco markers are attached to the robot to measure the $x$, $y$, and heading. The robot is remotely controlled via BLE. Our set-up is visualized in Figure ~\ref{fig:trainingsetup}. 

\subsubsection{Coarse Grid Search}
In our coarse grid search, users can control the maximum motor intensity (for both motor rotations), the number of intensities to visit per motor, and the step size. From our experimentation, we set the maximum motor intensity to 245, a step size of 40, and 2 intensities to visit. The maximum output voltage of our micro controller is 3.3V. Converting this into voltages, this filters the search space to all combinations containing -3.17V, -2.78V, 0V, 2.78V, and 3.17V for a total of 625 combinations. We chose these intensities from a series of prior experiments during the prototyping process where we discovered that low voltages fail to locomote the robot and high voltages rattle the robot uncontrollably. Users can select the number of pose samples per motor combination and the time difference between pose samples. In our case, we chose 25 samples with a 0.1 second spacing. The total duration of this approach for a robot takes 2 hours to complete including manual intervention for collision detection and battery charging for each side (front and back). The final output of the process is a .txt file that logs an input $m$ with a trajectory $P_{n}$ in the global coordinates. 

\subsubsection{Omni-directional and Rotational Motor Pattern Selection}
\removed{Based on initial results, our device is capable of locomoting omni-directionally through a singular motor pattern}

In order to identify specific motor intensities for omni-directional motion and rotation, we take the results of the training process and select the motor configuration that maximizes the function $\boldsymbol{R_i}$ for a target omni-directional heading $\hat\psi_i$. 
Trajectories are first transformed from global coordinates to the robot's relative coordinates.
We then take the arctangent, $\psi$ \removed{of}\added{using} the average x ($\bar{x}$), and y ($\bar{y}$) headings between the time samples. \added{The function is defined below:}
\removed{For target angle $\hat{\psi}$ and penalty $\lambda$ we define the below reward function.}
$$
\boldsymbol{R}_{i} =\begin{cases}
    max(\bar{x}_{i},\bar{y}_{i}) & \text{if } |\psi_{i} - \hat{\psi}_{i}| \leq \epsilon \\
    0  & \text{if } |\psi_{i} - \hat{\psi}_{i}| > \epsilon \\
\end{cases}
$$

Where $\epsilon$ is the error threshold. \added{The function selects motor configurations that yield in a discretized omni-direction with room for error.} In our approach, we use a threshold of $5^{\circ}$. After calculating the scores for all motor patterns and target angles, we obtained 36 motor intensities for omni-directional locomotion.

For rotations, from prior experiments, we filter the results of our training process to patterns where only one motor is actuated. From these options, we select the pattern that maximizes the difference in angular pose, $\theta$, and minimizes motor pattern intensity. This is done for both clockwise and counter-clockwise directions.

\subsection{Sensing}
\begin{figure}[h]
    \centering
\includegraphics[width=\figurewidth]{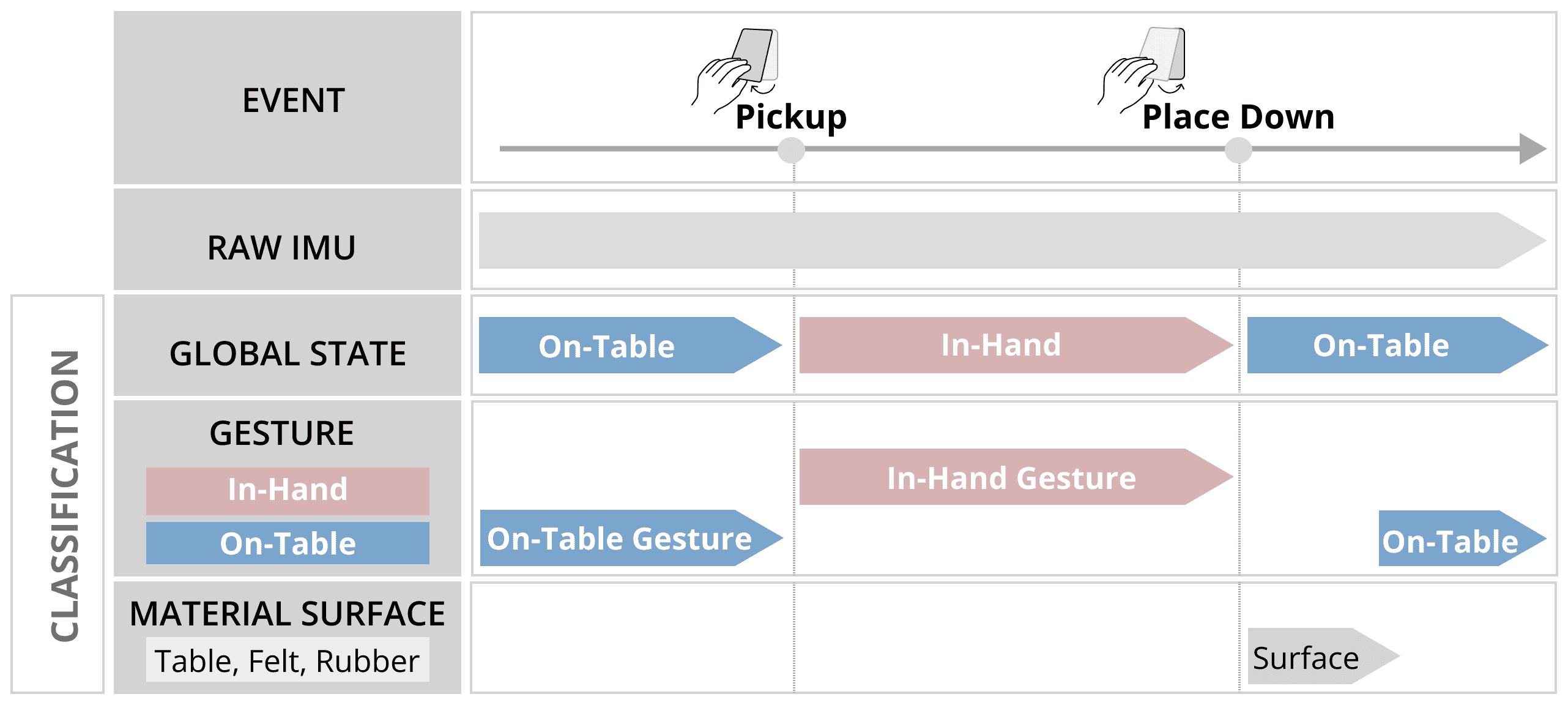}
    \caption{Sensing State Representation Diagram for Pickup and Place Down events, with regards to the four classification models.} 
    \label{fig:training-sensing}
\end{figure}

The sensing model is composed of 4 distinct models - global state classification, in-hand gesture classification, on-table gesture classification, and on-table surface texture classification (Figure \ref{fig:training-sensing}).

\begin{figure*}[ht]
    \centering
\includegraphics[width=\figurewidth]{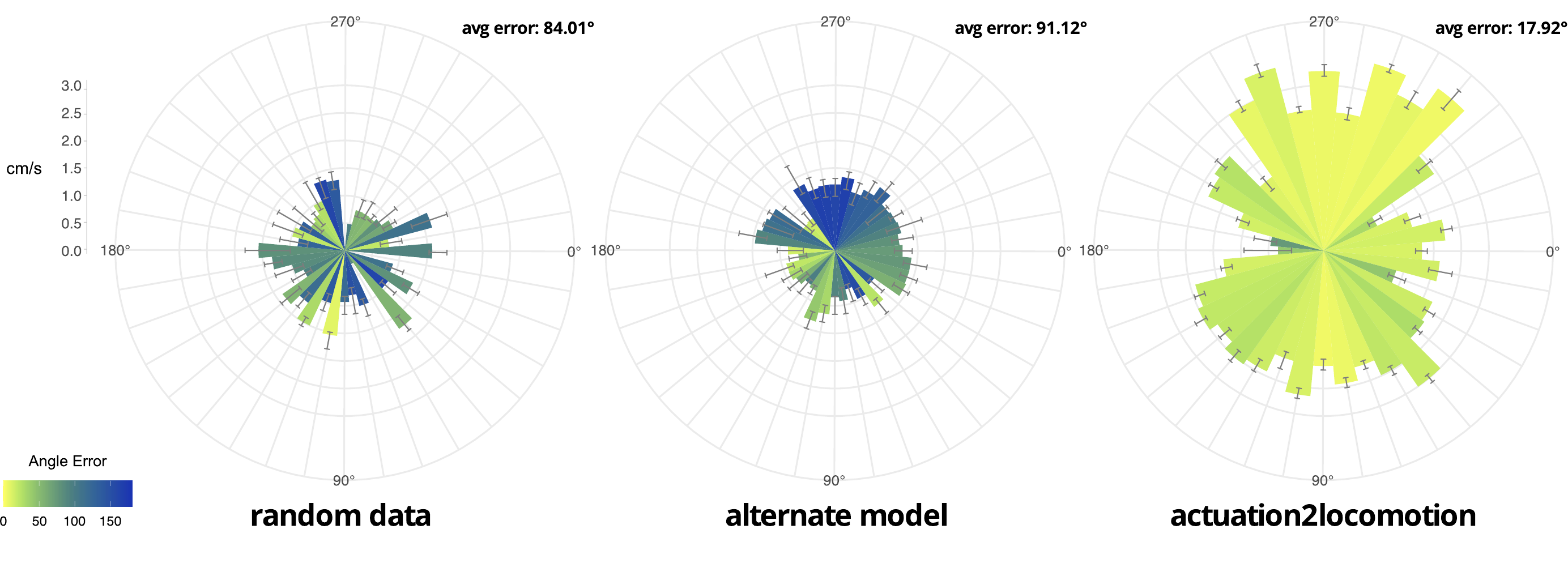}
    \caption{(Left) Average angle error and average velocity using a random sample of data. (Middle) Average angle error and average velocity using the \textit{actuation2locomotion} model from a different robot. (Right) Average angle error and average velocity post-training for a single device. \removed{Our model immensely improves the locomotion accuracy and the average velocity in the required direction.}}
    \label{fig:pre-post}
\end{figure*}

\subsubsection{Global State Classification}
\label{section:state-detection}
Classification is achieved using raw accelerometer data sampled at 416Hz to determine the current state of the device. A positive z-axis reading with near 0 $x$ and y-axis is classified as face-up on-table. Similarly, a negative z-axis reading with near 0 $x$ and the y-axis is classified as face-down on-table. Other inputs get classified as in-hand. 

\subsubsection{Gesture Classifications}
\label{section:Gesture Classifications}
We collected accelerometer data for gestures commonly used in card games. The accelerometer is sampled at 416Hz and aggregated by taking the max of the 3-axes. 200 frames are collected and sent via BLE for model creation. Taps and slides were collected for on-table interactions. Shakes, and flicks were collected for in-hand interactions. We divide the frame into 20 windows and perform feature extraction by computing the mean, standard deviation, min, and max. We utilize this to train a \added{one-vs-rest} logistic regression model.

\subsubsection{Surface Texture Classification}
\label{section:surface-detection}
For surface classification, we upsampled the IMU to 3.332Khz. 
Laput et. al has shown that upsampled accelerometer data on wrist-worn devices enables the classification of bioacoustic signals \cite{laput_viband_2016}. We deploy a similar approach for IMU preprocessing. \removed{The device is locomoted using a preset lower intensity for 1 second while simultaneously calculating IMU data.} \added{The device is actuated with the bottom left motor at 70\% intensity while IMU data is collected.} 64-point FFT is performed for each axis and then aggregated by taking the max of each frequency to create the input for the models. Once classified, we reset the IMU to its default settings. Textures we collected data for span various potential playing surfaces with varying textures and softness - \textit{felt}\footnote{\url{https://www.amazon.com/gp/product/B07CTQQLRP}}, \textit{card playing mat fabric}\footnote{\url{https://www.amazon.com/gp/product/B09R7VH4SX}}, \textit{frictionalized rubber}\footnote{back side of \textit{card playing mat fabric}}, and a laminate covered wooden \textit{table}. \added{100 samples of the IMU data is collected across the various materials} This collection is done across multiple devices to form our dataset. We trained a \added{one-vs-rest} logistic regression model on the FFT bins withholding 20\% of the data as a test set. This model was deployed on board and can be done when the robot is placed face-up on the table.

 \section{Technical Evaluation}
\label{section:tech-eval}

In this section, we first verify the functionality of the training process and that each device requires a unique \textit{action2locomotion} model.
Vibrational locomotion is inherently stochastic, and understanding its \textit{robustness} and \textit{transferability} is important to designing generalizable applications. We evaluate both the robustness and the transferability of the training process described in \removed{Section 5}\added{Section~\ref{section:training}}. In the first set of experiments, we evaluate the robustness by testing the reproducibility of the \textit{actuation2locomotion} model across 3 devices. In the second set of experiments, we test the transferability of the \textit{actuation2locomotion} model. During the training process, we use a single standard surface to build the \textit{actuation2locomotion} model. As our system is designed with customization in mind, testing the transferability of the model across customizations and surfaces allows us to understand the limitations of the training setup and to evaluate whether re-training the model is necessary to accommodate different conditions. Additionally, we evaluate the performance of surface detection of our robot. Gesture will not be evaluated as classification using IMU is well researched~\cite{wu2009gesture}.

\subsection{Training Process Verification}

Before we understand the robustness of our model across devices, we first need to verify its effect on a single device. For a single front-facing device, we run the training procedure outlined in Section~\ref{section:training} and extract the \textit{actuation2locomotion} model. In this experiment, we use the same robot to evaluate performance. We ran the 36 motor patterns 10 times to evaluate the accuracy of the locomotion trajectory. \removed{This process is visualized in} \removed{Figure~\ref{fig:omni_selector} \added{despicts path of the robot following the training process, selection of omni-directional trajectories, and the repeatability experiments}.} Using the full training data, first, we randomly assign a motor pattern to a discretized direction and run the evaluation. Next, we utilize the \textit{actuation2locomotion} model from a different device to ensure that models are not transferable across devices. For the last condition, we utilize the specific device's \textit{action2locomotion} model. Our results are summarized in Figure~\ref{fig:pre-post}. We compute the average angular error and velocity for each $10^\circ$ segment. We find that a \textit{personalized} action2locomotion model works best and surprisingly, a transferred model from a different device performs worse than randomized data. 

\subsection{Training Robustness}
\begin{figure}[h]
    \centering
\includegraphics[width=\figurewidth]{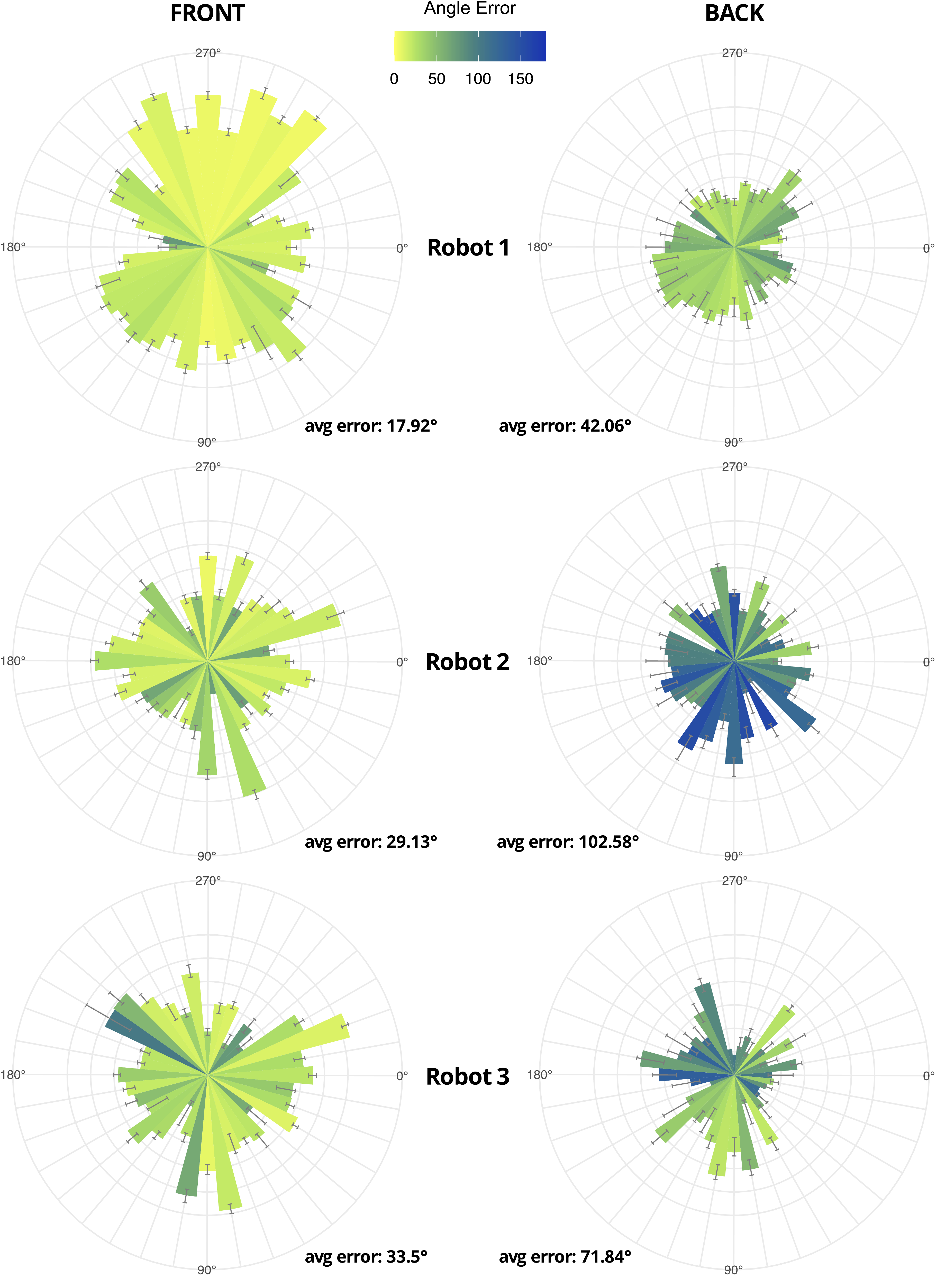}
    \caption{Visualizing average angle error and average velocity across 3 devices (front and back)}
\label{fig:trainingrepeatability}
\end{figure}

Using results from the prior section, we ran individualized training for 3 devices on both sides (front and back). We ran the 36 motor patterns 10 times and re-evaluated the accuracy of the locomotion trajectory. We visualize the results in Figure ~\ref{fig:trainingrepeatability}.

We find that our aggregated average error of the robot's trajectory is 26\%. Due to the scale of how the data is collected, small perturbations (1mm) can greatly impact the results. When looking at individual target angles and paths, a majority of our discretized motor patterns have low angular errors. The preliminary results look promising to locomote omni-directionally face-up\added{, especially since i}\removed{. I}n card and board games \removed{precise targeting is not important and }locations \added{mainly} exist in zones. 

Evaluating the back-side of the card, our aggregated average error is 72\%. \removed{Literature in legged robotics is vast and these results align with prior work} The addition of legs bias locomotion to certain directions and thus limits omni-directionality. For practical reasons, we employ this design decision - however, these results are promising in our exploration of leg-less flat robotics.

The results also seem to suggest that there's a degree of variance in the robustness between each device. While this can be an artifact of the stochastic nature of vibration-based locomotion, additional inquiry can be directed towards improving the accuracy. Our architecture is novel, and the lack of a baseline creates challenges in evaluating our approach. Should users prioritize accuracy, a traditional rotate-and-go approach used by differential-driven robots may be more suitable instead of an omni-directional approach.

\subsection{Transferability}
Using the base \textit{action2locomotion} model, we evaluate its transferability \removed{to new environments and modifications}. \added{Transferability is defined as 
"reusing previously learned parameters in unseen scenarios" \cite{jaquier2024transferlearningroboticsupcoming}}. \added{In our context this would be new surfaces and card customization.}
removed{Specifically surfaces that differ from our initial training process and customization that alter the contact point between the robot and the surface.} Based on the locomotion characteristics observed, the \textit{friction} and \textit{impact} between the device and the surface contribute to the performance of the locomotion. We define transferability by comparing the results of the robustness tests with the below experiments. Since the training process is cumbersome, we evaluate this to minimize the amount of retraining time for users if they were to make modifications or play on different surfaces. Results that yield similar angular trajectories at a lower/higher magnitude indicate that frictional forces are linearly applied and thus these conditions are considered transferable. If the results completely alter the baseline robustness tests, we recommend retraining the device on that unique alteration. In this section, we reduce the granularity of omni-directional from 36 to 18 and run 5 samples for each instead of 10. Furthermore, evaluation is prioritized on the face-up side of the device as omni-directional locomotion in this direction is more robust. 

\subsubsection{Surface Textures}
We identified three common playing surfaces for table-top interaction - felt, frictionalized rubber, and a card playing mat (details of these materials can be found in 5.2.3). We select distinct textures and softness to evaluate how they may potentially affect locomotion. 

\begin{figure}[h]
    \centering
\includegraphics[width=\figurewidth]{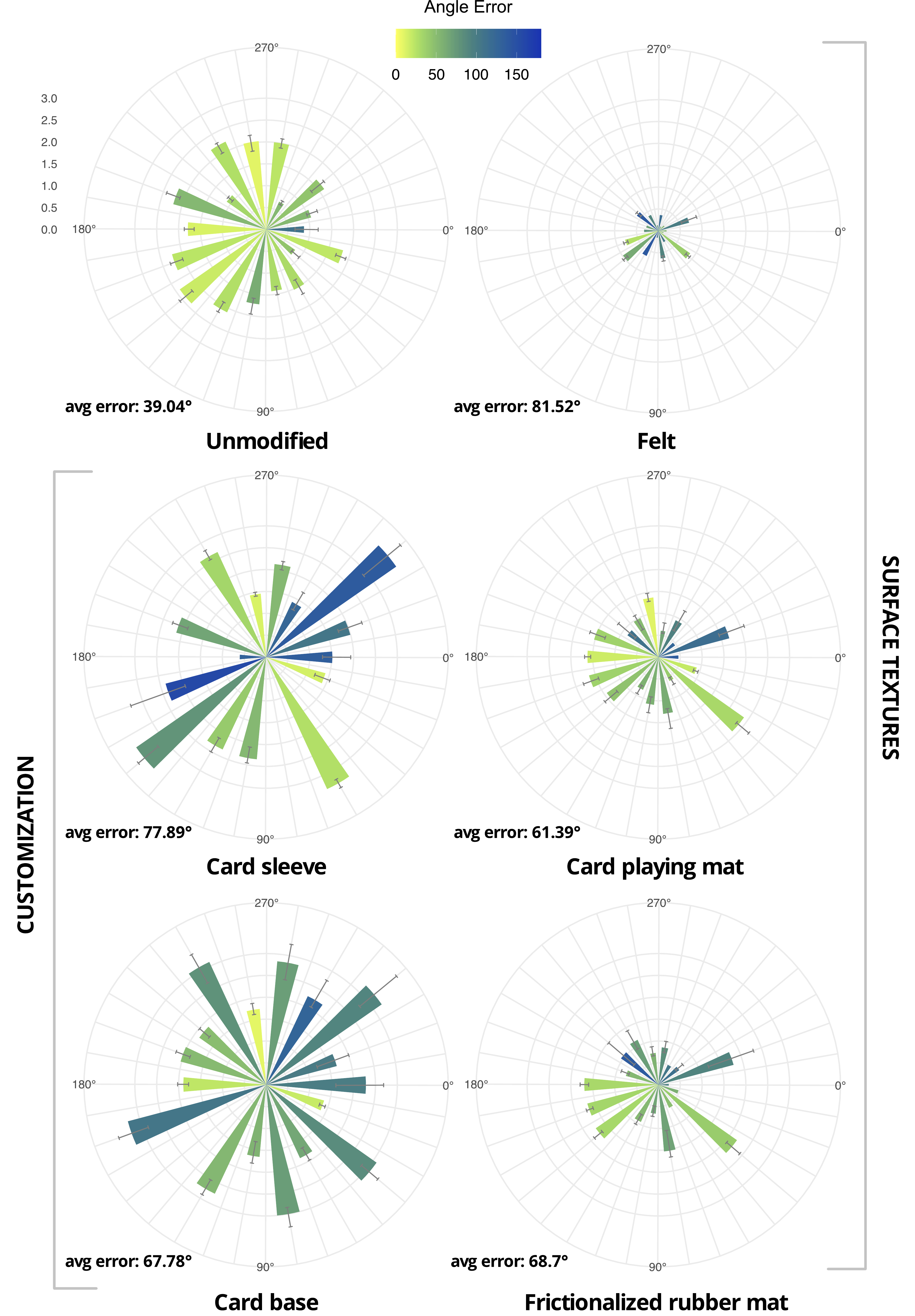}
    \caption{ Top left shows the error plot for an unmodified device. (Left) Error Plots for Evaluation of Customization, including Card Sleeve and Card Base (Right) Error Plots for Evaluation of Surface Texture, including Felt, Card Playing Mat, and Frictionalized Rubber Mat.}
    \label{fig:transfer}
\end{figure}

Summarizing the results from Figure~\ref{fig:transfer}, we can conclude that while felt significantly dampens the motion, it is sufficiently accurate in certain directions. The card playing mat and the frictionalized rubber mat don't dampen the movement as much as the felt does and is more accurate when compared to the felt. 

\subsubsection{Robot Customizability}

The affordances of the card shape enable users to customize the card. Customization alters the friction between the device and the surface. We use commonly available Bicycle playing cards\footnote{\url{https://bicyclecards.com/}} with tactile patterns as the \textit{card base} and a non-slip playing \textit{card sleeve}. As seen in Figure~\ref{fig:transfer}, customizations to the device reduce the friction between the robot and the surface, resulting in dramatically higher average velocities but the accuracy is significantly lower when compared to the unmodified device. This suggests that retraining is necessary when modifications are made to the contact surface between the robot and the surface.

\subsection{Surface Texture Classification}

As discussed in Section~\ref{section:surface-detection}, we deploy a sensing model on board and collect 100 classification samples for each surface texture. The results plotted below are the output of the hold-out set when training the robot's on-board model.

\begin{figure}[h]
   \centering
\includegraphics[width=\figurewidth]{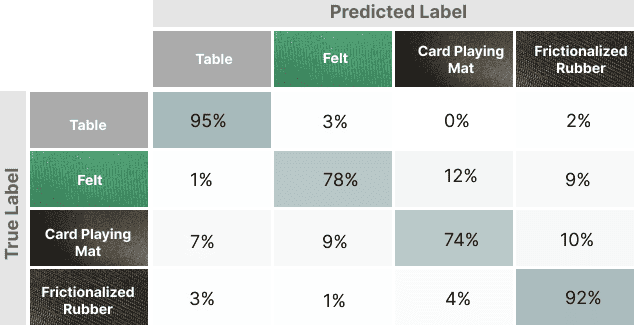}
   \caption{Confusion Matrix for Surface Texture Classification}
\label{fig:confusion_matrix}
\end{figure}

Enabling the robot to detect the surface on which it locomotes on, creates additional interactive opportunities. These opportunities are explored in the application section.
  \section{Applications}
\label{section:applications}

\system{}'s interaction capabilities grounded in the vibration-oriented actuation for haptics and locomotion open up a wide range of applications based on the affordance and utility of cards. We demonstrate applications across card games for entertainment, flashcards for learning, and other card-based utilities.

\subsection{Card Games}

Card games are one of the major applications of \system{}. We demonstrate card game applications in two potential directions: \textit{1) how \system{} can augment and guide existing\removed{card} games}, and \textit{2) how \system{}'s capabilities can be employed for new game mechanics}.

\begin{figure}[t]
    \centering
\includegraphics[width=\figurewidth]{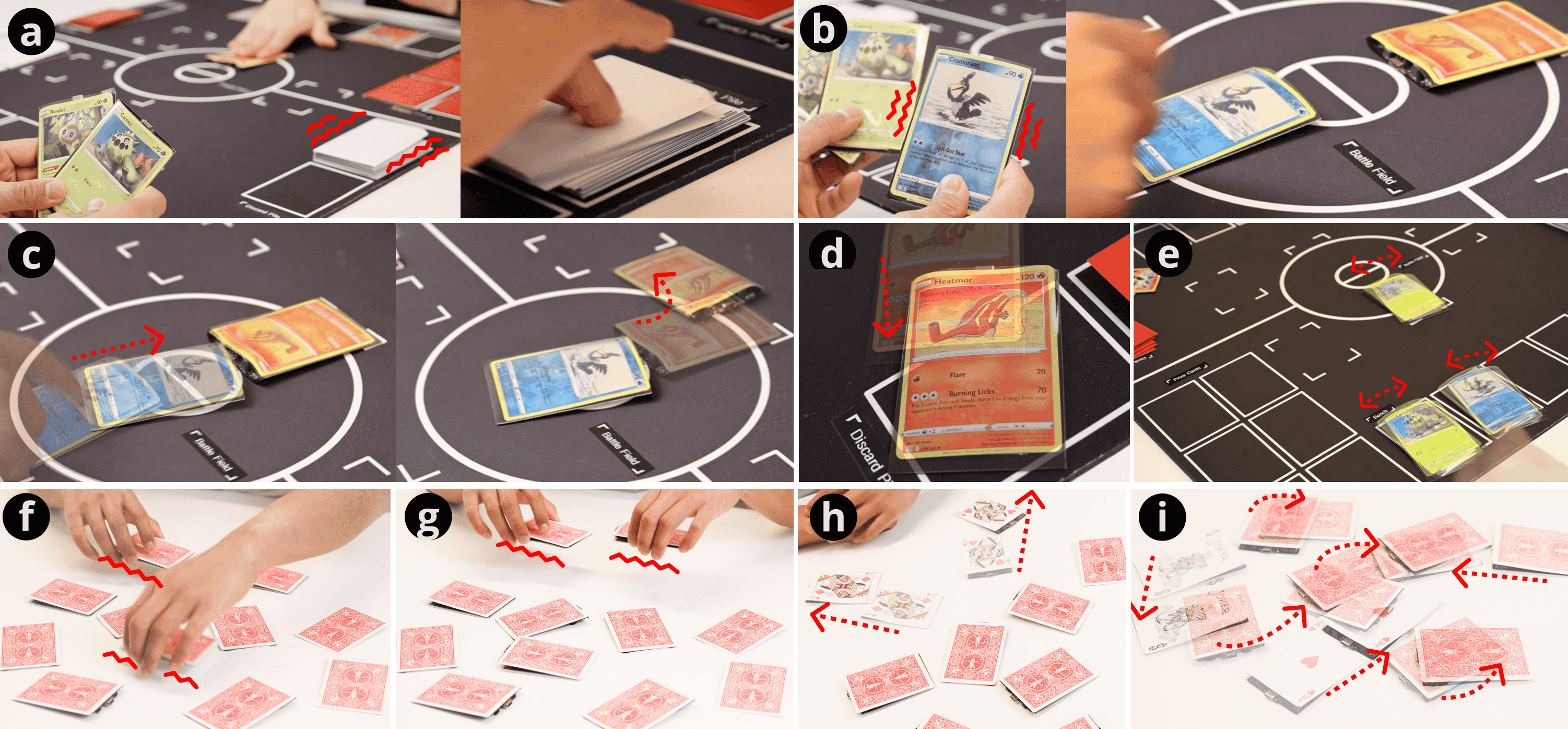}
    \caption{Monster Dueling Card Game Application \added{(a-e) and Haptic Matching Card Game (f-i)}\removed{, guiding users to play through haptics and buzz-signal (a, b), and using locomotion capabilities to enhance the expressibility of the card characters (c-e).}}
    \label{fig:battling game}
\end{figure}

We demonstrate the first direction through \textbf{a monster dueling card game}, which is one of the most commercially successful types of card games, as found in different brands, Pokemon\footnote{\url{https://www.pokemon.com/us/pokemon-tcg}}, Yu-Gi-Oh!\footnote{\url{https://www.yugioh-card.com/}}, or Magic the Gathering\footnote{\url{https://magic.wizards.com/}}. These games are commonly played between two players who `summon' monsters to battle each other \added {by picking up a card from the deck and strategically choosing a character to battle on the field}. In such a game, \system{} could guide users to play, which might be useful especially for beginners. As shown in Figure \ref{fig:battling game}, the \removed{buzz signal}\added{Buzz Notification}\removed{on the top of the deck} could notify and remind the player to pick up a card every time their `turn' starts (a).\removed{Once it is their turn} To summon a character, subtle vibration in the hand can provide secret suggestions to the users about which character \removed{they should better place in the field to play advantageous}\added{to play} (b). 

Additionally, the locomotion capability of\removed{the} \system{} would contribute to adding expressability to the card characters, which greatly enriches the\removed{entertainment/} storytelling potential of these\removed{card} games\removed{via actuation}. For example, \added{as compared to a traditional manual placement of the cards,} the battling scene could be enhanced through the \removed{actual}\added{autonomous} motion of the card to `stage' the battle, making a \removed{n actual}\added{physical} collision between two cards (c), and when one character loses, \removed{they can}rotat\removed{e}\added{ion} to express `lay-down' and \removed{then move}\added{moving} to the `Discard Pile' \removed{autonomously}(d). If a player wins, all of their character cards on the playing mat can sync to rotate left and right, expressing `dancing' to celebrate their victory (e). 

The \system{}'s capabilities also have great potential to introduce unique game mechanics into card games. Figure \ref{fig:battling game} represents \textbf{Haptic Card Matching Game},\removed{ as} a simple example of such direction. 

In this game, a player uses each of their hand to lift two cards simultaneously, which would activate the cards to play certain vibration patterns. When they feel different haptic patterns on both of them (f), they have to place them down and pick another pair to find a matching pair (g). Once they find a match, they flip the card to confirm via the graphical pattern on the other side of the cards, while these cards move by themselves to be obtained by the player (h). Additionally, this game can incorporate a raffling feature: a Joker card acts as making a fake haptic signal, that when the player flips it, all the cards on the table start moving to shuffle themselves (i), making the matching process challenging.

\subsection{Actuated Flashcards}

\begin{figure}[t]
    \centering
\includegraphics[width=\figurewidth]{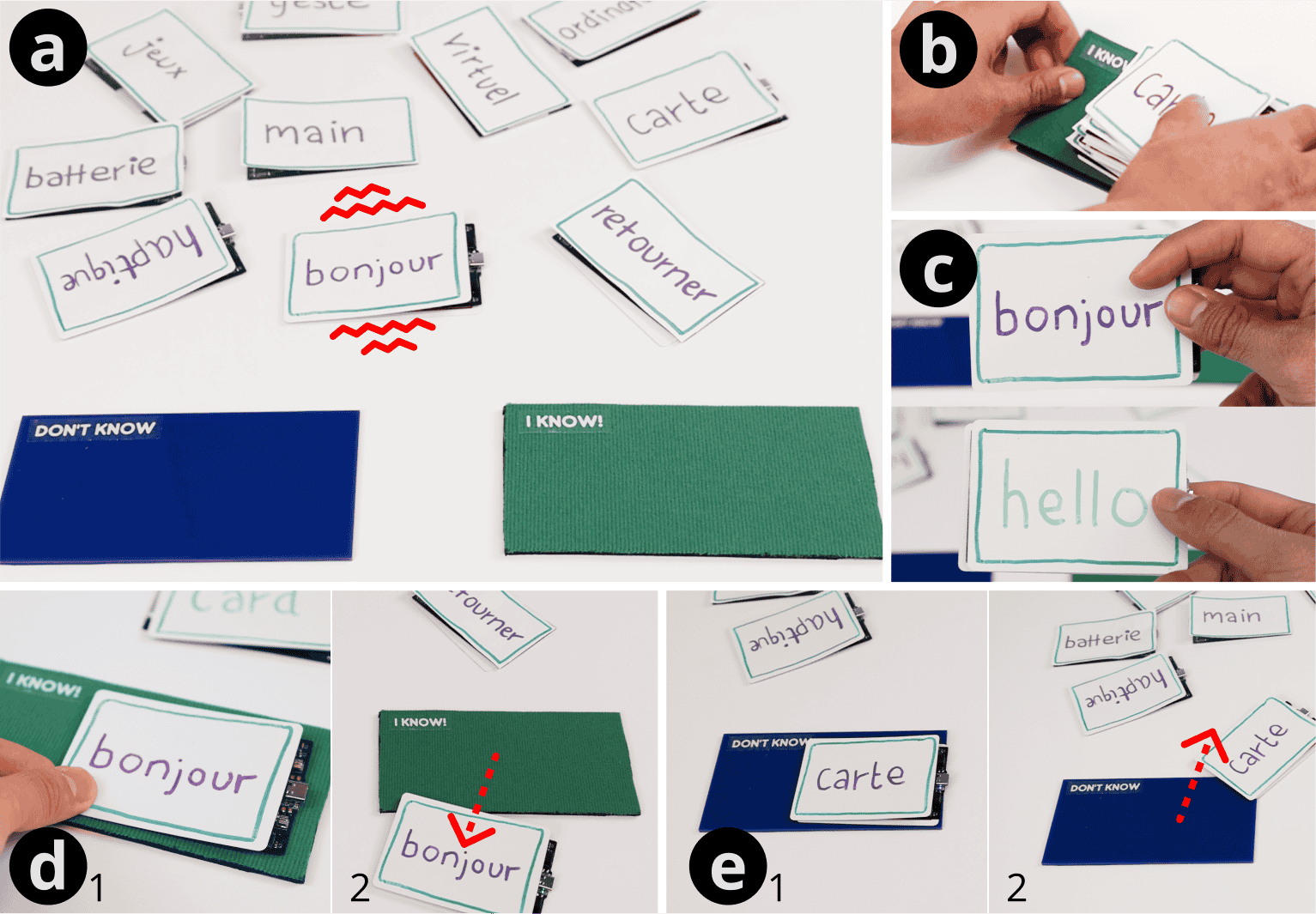}
    \caption{Flashcard Application for vocabulary learning \removed{, the card can signal to be picked up (a), recognize if it is placed on the `know' or `don't know' surface and move accordingly (d,e).}}
    \label{fig:educational game}
\end{figure}

\system{} has great potential as a learning tool, as cards are often used for educational materials. Figure \ref{fig:educational game} shows a flashcard application\removed{ of \system{},} where \removed{the} French vocabulary is written on one side and corresponding English words on the other (a, c). The entire kit comprises a set of cards and two small mats \removed{made} of different materials that can be compactly stored (a, b). One of the cards on the table nudges the user to pick it up by vibrating (a). If they remember the word, they may place the card on the `I KNOW!' mat, made of felt (d1), where the device detects the \removed{materiality}\added{surface }of the mat and moves to the bottom side of the table \removed{surface}(d2),\removed{as} a zone\removed{of} \added{for} `already learned.' On the contrary, when a card is placed on the `I DON'T KNOW' mat, made of acrylic (e1), the device can move upwards\removed{where} \added{with the} other cards that have not been learned\removed{ are placed} (e2).

While physical flashcards are preferred for a certain group of people due to tangibility and spatial memory \cite{nakata2019learning}, they could incorporate benefits from digital flashcards (e.g. ones that can be found on browser or smartphone apps), which, for example, could track the words the user need to repeatedly learn. Such a system could also keep track of the vocabulary that has not been picked up to actively support users learning through actuation and tangibility.

\subsection{Everyday Cards}

\removed{Beyond card-based games and education tools, we share how \system{} could enrich other everyday card-based interactions.}

\begin{figure}[t]
    \centering
\includegraphics[width=\figurewidth]{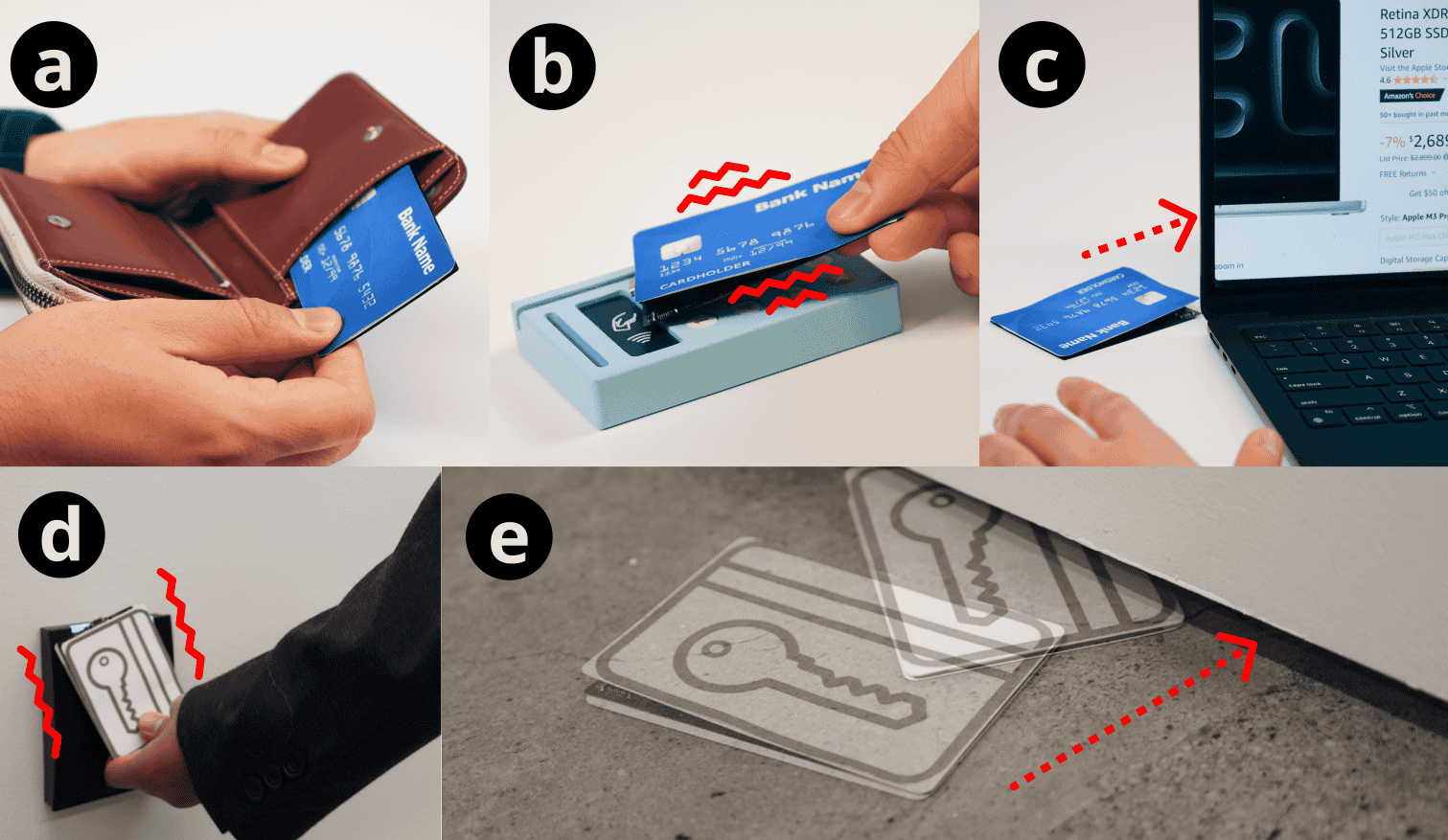}
    \caption{Credit Card Application \added{(a-c) and Card Key Application (d,e)}\removed{, fitting into wallet (a), vibration as haptic confirmation of payment (b), and the card running away from the user to recommend spending less money (c).}}
    \label{fig:credit card}
\end{figure}

\added{We share how \system{} could enrich everyday card-based interactions.} The thinness of the device allows it to fit into a wallet, for example, as a \textbf{credit card} (Figure \ref{fig:credit card} a). When \system{}'s technology is incorporated into credit cards, the vibration could provide haptic confirmation for the payment, providing a tangible feeling for an intangible transaction (b). As credit cards' NFC-based touch payments are raising security concerns \cite{akinyokun2017security}, fusing haptic feedback into the cards may help the cardholders notice an unintended payment effectively. Additionally, as the balance and usage history for the credit card is often hard to keep track of, the credit card could\removed{abruptly} actuate to nudge the user not to spend beyond what they wish to, for example, by locomoting away from the user when they are browsing online for an impulsive expensive purchase (c).

Card keys are widely used as a thin mobile form of security access with NFC readers for hotels, schools, and offices. \system{}'s vibration capability may provide haptics to indicate the lock status of\removed{the} doors for intuitive interaction (Figure \ref{fig:credit card} d), similar to the credit card idea. \removed{Furthermore, t}\added{T}he locomotion capability \added{also} could make the card key move by itself when left inside a locked door. Specifically, the thinness of the device (< 6mm) makes it possible to move the key from under doors (e). As there are some card-shaped IoT tracker products in the market\footnote{\url{https://www.tile.com/product/black-slim}} -- helping the user to find the tracker by beeping or sharing GPS information --, we believe making the card itself moves is a future direction such a product could take.

 \section{Discussion and Future Work}
\label{section:discussion}
The paper introduced a new form of interactive robots, which take a flat, thin, card-like shape to introduce novel interaction and application. While our hardware implementation and prototypes opened up unique interactivity, there are limitations and future directions that are \removed{essentially}exciting future research opportunities\removed{, which we would like to explicitly share}. 

\subsection{Hardware Limitation and Update}

\textbf{Noise:} One \removed{of the obvious} limitation\removed{s} of our approach is the \textbf{noise of the vibration}, which can be a great disadvantage for some user scenarios \removed{,}\added{such as} \removed{for example,} game designs that require quiet narrative and game mechanics. Vibration can \removed{simply}be mentally stressful to constantly hear\removed{for certain groups of people}. Reduction of noise \removed{, or building an environment that the noise won't be distinct (e.g. noise cancelling) could be further explored.} \added{could be achieved by investigating sound-absorbing materials on the card or surface. The mitigation of the sound with white noise or noise cancelling environments could also be explored.}

\textbf{Towards Complete Flat-Shape:} While the main body of our current prototype is thin and semi-flexible \removed{thanks to flexible PCB and thin LiPo battery,} aproximately 22 \% percent of the surface area is composed of thick, rigid components, that could be replaced in the future towards \textbf{completely flat, thin, semi-flexible form}. \removed{Among them, vibration motors are the bottle-necks to keep the device thin}\added{ERMs limit the thiness of the device}. \removed{While we excluded them from our implementation to make large locomotion and haptics,o}\added{O}ther vibration actuators (piezo-film, LRA vibration motors, etc.) could be explored to make the entire body flat and thin. \added{This work prioritizes large locomotion and haptics}

\textbf{Adding Display:} Exploring \textbf{extended I/Os} would be another avenue to enrich the device's interactivity. While, for our prototype's device appearances, we have simply attached static printed images, future systems could incorporate \textbf{thin, flexible display technologies} to dynamically change the appearance of the device, to convey different card-based visuals. A system like the e-ink display could match the concept of the device, that keeps the appearance and materiality of cards, rather than illuminating displays.

\textbf{Extended Sensing:} While the actuation was the core focus of our paper, \textbf{advanced sensing} to track interactions unique to cards' affordances is an exciting technical challenge\removed{s}. Such sensing could detect how cards are being touched, pressed, or bent by users. While we demonstrated camera-based tracking for closed-loop targeting, it would be best if the robots could self-identify their position without external tracking systems. This could be enabled by taking on-board tracker, such as the computer vision method introduced in toio robots \cite{toio}.\added{Additionally robots that can detect their position to other robots can open up interaction possibilities of the system} \removed{Furthermore, informally talking to a board game designer in our institution raised different sensing requirements and possibilities, such as cards being able to detect their relative position for other cards. Further advancing the I/O based on potential users of the system is one of the ideal research approaches to build a technical system, matching user needs.}

\removed{\textbf{Other Actuation:} Future \system{} could incorporate \textbf{other types of physical actuation}, such as shape-changing \cite{heibeck2015unimorph, roudaut2013morphees}, thickness-changing, or weight-shifting. Incorporation of these actuation methods needs to be carefully considered based on the trade-off of keeping the device form factor, especially the thinness and battery capacity. \quad{I think this is unrelated/lower priority}}

\begin{figure}[h]
    \centering
\includegraphics[width=\figurewidth]{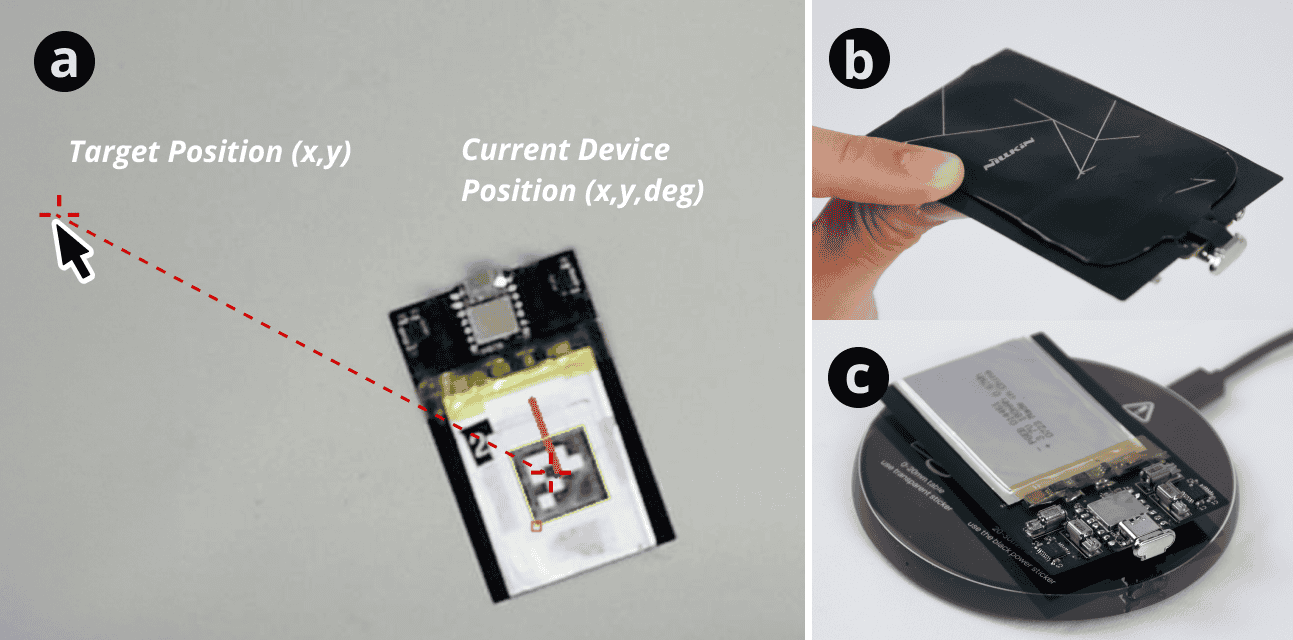}
    \caption{\added{Computer vision-based closed-loop control GUI system (a),} addition of Wireless Charging Pad (1.7mm) to \removed{Cardinality} \added{CARDinality} Robot (b), and the Robot being charged with a charging pad over plexiglass with thickness of 5mm (c) .}
    \label{fig:wirelesscharging}
\end{figure}

\textbf{Battery and Wireless Charging:} \removed{Power is another technical challenge to be tackled in the future.} In our prototype, the battery lasted for 35 mins, when all the motors were actively actuated and at maximum intensity. \removed{(For our}\added{During} training, it lasted \removed{around} 70 mins, and for standby mode, it lasted more two days\removed{)}. This battery capacity is not ideal for card games or board games\removed{, that can last much longer}. This problem could potentially be addressed using a wireless charger. Using our prototype, we have verified that the device was able to be charged using an off-the-shelf USB-C wireless charging receiver pad (1.7mm thick) (Figure \ref{fig:wirelesscharging}). Future systems should incorporate a method for the device to track charging stations, to self-locomote onto it when the battery is low. 

\textbf{Device Robustness:} We have noticed from our hardware that, \textbf{hardware robustness} is one thing that further needs to be evaluated\removed{, and assessed}. Specifically\removed{ we have noticed} vibration sometimes breaks the soldered battery connection, or glued 3D printed casing \removed{(though in most of the training process, they rarely break.)}\added{.} Improving the robustness by mitigating the vibration effect is important in the future, for example by applying material that can absorb vibration energies. 

\subsection{Training Approach for Locomotion}

The limitation of our system is the cost to sample motor patterns and observe their output trajectory. While the results from our coarse grid search approach are promising, there is no guarantee that optimal motor configurations occur within our constraint and thus it is beneficial to expand the search space. Furthermore, our approach personalizes a locomotion model for each robot to account for potential robot deformations in the fabricating process which \removed{significantly} increases the training time for swarm-based interactions. The lack of an analytical model \removed{constrains efficiently scaling up}\added{limits efficient scaling of} the search space of the training process\removed{. An analytical model of our robot would} \added{by} enabl\removed{e}\added{ing} simulation-based sampling and learning. Transferring \textit{controllable} simulation-learned parameters to real-world locomotion has previously been proposed to improve the efficiency of robotic training approaches \cite{tan2018Sim2Real} to speed up the training process. \added{Additionally, the current training approach requires constant supervision. We can improve this process by creating a mechanism to recenter the device.}
\removed{
Additionally, the current training approach requires constant supervision. We can improve this process by creating a mechanism to recenter the device.}\removed{ This can enable users to parallelize the approach and collect data for multiple robots simultaneously.}

\subsection{Extended States, Utility and Interactivity}

\removed{The applications and utility presented in the paper are only part of the possibility of \system{}, and we should further explore extended utility,interactivity and applications.}

While our paper focused on On-Table and In-Hand states, \removed{the two major states cards are commonly used in, there are} \textbf{Other States} \removed{the cards are being used in everyday environments}\added{such as In-Deck, In-Pocket, and In-Wallet could be further explored from a sensing and software system perspective.} \removed{, including in-wallet, in-pocket, in-stack, etc. By considering these states, the system could track and detect these states to adaptively and interactively provide actuation feedback.}\added{Also, as cards are often}\removed{As there are many instances cards are being} `inserted' into devices\removed{, tools,} and materials, \removed{we consider that could be an interesting research opportunity}\added{it would be interesting} to use the card to activate \removed{other} passive objects by inserting and propagating \removed{the}vibration-based actuation, as in \added{Hermits}\cite{nakagaki2020hermits}. 

One unique functionality we found in the final stage of the prototype is that\removed{,} the vibrating card combined with \removed{slippery card}\added{specific} material\added{s} can greatly affect \textbf{how \removed{much}\added{smoothly} the card \removed{slips and }slides over\removed{smooth} tabletop surfaces}\footnote{Please watch the end of supplementary video for the reference.}. \removed{In such a case, v}\added{Similarly to T-PaD~\cite{Winfield2007tpad}, which uses variable friction reduction for haptics using vibration, CARDinality's v}ibration can be \removed{further}controlled to dynamically tune the slipperiness of the card, \removed{that}\added{which} could provide new game mechanics and utility. 

\added{As an early exploration of further interactivity with CARDinality, we also constructed a computer vision-based closed-loop control system for the GUI application. When launching this feature, users can click on the camera view window to provide a specific target (x,y) position for the robot (Figure \ref{fig:wirelesscharging}a). With the use of Aruco markers, the computer vision tracks the current device position (x,y,deg). This is used to calculate a vector between the robot and target to select the discrete omni-directional movement the robot should perform. The robot actuates using the corresponding pre-trained motor patterns. While this is only a preliminary interface, it opens up a variety of more fine-tuned locomotive applications for CARDinality.}

\removed{To further explore applications, we are extremely interested in conducting hands-on workshops to prototype and ideate a broad set of applications. Specifically, we are planning to host \textbf{Game Jam Sessions}, a common activity for game designer community to build and prototype games (similar to hack-a-thon for engineers). While the game design examples in the application section were by authors (no experience in game design), we believe gaining extensive inputs from experienced game designers would provide more unique ideas that fully incorporate what \system{} can do. \quad{Since are making a poster, can we potentially remove this}}

\subsection{User Study, and User Reaction} 
In our private research prototype exhibit event, visitors \removed{have} reacted that our latest prototype, indeed, feels like \added{a} card\removed{s}, allowing them to apply many conventional card-based affordances.\removed{ Such aspects would be interesting to study through thorough evaluation methods}. Although our paper focused on hardware development, design space, and application exploration, \removed{we would like to conduct further}user studies \added{could help us}
\removed{to} understand how people interact with the devices. Some major research questions include ``How would people interact with the \system{} device, in terms of affordances?'' ``How would the vibration noise affect the perception and interaction with the device?'' ``How would people interpret different haptic patterns and locomotion modes?'' To answer these questions, carefully crafted empirical and quantitative study designs is required, together with a reproducible device design. 

 \section{Conclusion}
\label{section:conclusion}
In this paper, we introduced \removed{Cardinality}\added{CARDinality}, a novel interactive mobile robots, leveraging the form factor of cards. The robot is equipped with vibration-based actuation capabilities to serve both locomotion and haptic feedback and contains wireless control, IMU-based sensing, and a thin LiPo battery. Through this implementation, we have opened up a novel interaction design space that leverages the affordance and utility of card-shaped objects fused with actuation. As \removed{vibration-based locomotion} \added{vibration-based omni-directional sliding locomotion} and haptics make the actuator possible to be encapsulated, the robot can be customized by inserting in card sleeves and taping cards and pieces of paper. We presented a variety of applications grounded in card-based interaction, from gaming and learning to other everyday card usage scenarios. The applications, together with the technical evaluation, demonstrated the rich potential of the proposed robotic hardware, fusing robotic interaction capabilities into one of the common everyday tangible materials, Card.  
\begin{acks}
We thank Actuated Experience Lab members for their help with the project. The authors especially thank Ran Zhou for designing part of the diagrams in this paper. We thank Ramarko Bhattacharya for assisting in developing the GUI. We thank Lilith Yu, Steven Arellano, and Sophie Li for proofreading the paper. We thank Ashlyn Sparrow for brainstorming applications for the game design community.  Lastly, we acknowledge Togay Atmaca, who originated the research direction in this paper. 
\end{acks} 
\bibliographystyle{ACM-Reference-Format}
\bibliography{sample-base}


\begin{thebibliography}{59}


\ifx \showCODEN    \undefined \def \showCODEN     #1{\unskip}     \fi
\ifx \showDOI      \undefined \def \showDOI       #1{#1}\fi
\ifx \showISBNx    \undefined \def \showISBNx     #1{\unskip}     \fi
\ifx \showISBNxiii \undefined \def \showISBNxiii  #1{\unskip}     \fi
\ifx \showISSN     \undefined \def \showISSN      #1{\unskip}     \fi
\ifx \showLCCN     \undefined \def \showLCCN      #1{\unskip}     \fi
\ifx \shownote     \undefined \def \shownote      #1{#1}          \fi
\ifx \showarticletitle \undefined \def \showarticletitle #1{#1}   \fi
\ifx \showURL      \undefined \def \showURL       {\relax}        \fi
\providecommand\bibfield[2]{#2}
\providecommand\bibinfo[2]{#2}
\providecommand\natexlab[1]{#1}
\providecommand\showeprint[2][]{arXiv:#2}

\bibitem[noa({[n.\,d.]})]%
        {noauthor_method_nodate}
 \bibinfo{year}{[n.\,d.]}\natexlab{}.
\newblock \bibinfo{title}{Method {Cards} — ideo.com}.
\newblock
\newblock
\urldef\tempurl%
\url{https://www.ideo.com/journal/method-cards}
\showURL{%
\tempurl}


\bibitem[Akinyokun and Teague(2017)]%
        {akinyokun2017security}
\bibfield{author}{\bibinfo{person}{Nicholas Akinyokun} {and}
  \bibinfo{person}{Vanessa Teague}.} \bibinfo{year}{2017}\natexlab{}.
\newblock \showarticletitle{Security and privacy implications of NFC-enabled
  contactless payment systems}. In \bibinfo{booktitle}{\emph{Proceedings of the
  12th international conference on availability, reliability and security}}.
  \bibinfo{pages}{1--10}.
\newblock


\bibitem[Altice(2014)]%
        {altice_playing_2014}
\bibfield{author}{\bibinfo{person}{Nathan Altice}.}
  \bibinfo{year}{2014}\natexlab{}.
\newblock \showarticletitle{The playing card platform}.
\newblock \bibinfo{journal}{\emph{Analog Game Studies}} \bibinfo{volume}{1},
  \bibinfo{number}{4} (\bibinfo{year}{2014}).
\newblock


\bibitem[Alves and Roque(2011)]%
        {alves_deck_2011}
\bibfield{author}{\bibinfo{person}{Valter Alves} {and} \bibinfo{person}{Licinio
  Roque}.} \bibinfo{year}{2011}\natexlab{}.
\newblock \showarticletitle{A deck for sound design in games: enhancements
  based on a design exercise}. In \bibinfo{booktitle}{\emph{Proceedings of the
  8th {International} {Conference} on {Advances} in {Computer} {Entertainment}
  {Technology}}}. \bibinfo{publisher}{ACM}, \bibinfo{address}{Lisbon Portugal},
  \bibinfo{pages}{1--8}.
\newblock
\showISBNx{978-1-4503-0827-4}
\urldef\tempurl%
\url{https://doi.org/10.1145/2071423.2071465}
\showDOI{\tempurl}


\bibitem[Belke and Paik(2017)]%
        {belke2017mori}
\bibfield{author}{\bibinfo{person}{Christoph~H Belke} {and}
  \bibinfo{person}{Jamie Paik}.} \bibinfo{year}{2017}\natexlab{}.
\newblock \showarticletitle{Mori: a modular origami robot}.
\newblock \bibinfo{journal}{\emph{IEEE/ASME Transactions on Mechatronics}}
  \bibinfo{volume}{22}, \bibinfo{number}{5} (\bibinfo{year}{2017}),
  \bibinfo{pages}{2153--2164}.
\newblock


\bibitem[Caraban et~al\mbox{.}(2020)]%
        {caraban_nudge_2020}
\bibfield{author}{\bibinfo{person}{Ana Caraban}, \bibinfo{person}{Loukas
  Konstantinou}, {and} \bibinfo{person}{Evangelos Karapanos}.}
  \bibinfo{year}{2020}\natexlab{}.
\newblock \showarticletitle{The {Nudge} {Deck}: {A} {Design} {Support} {Tool}
  for {Technology}-{Mediated} {Nudging}}. In
  \bibinfo{booktitle}{\emph{Proceedings of the 2020 {ACM} {Designing}
  {Interactive} {Systems} {Conference}}}. \bibinfo{publisher}{ACM},
  \bibinfo{address}{Eindhoven Netherlands}, \bibinfo{pages}{395--406}.
\newblock
\showISBNx{978-1-4503-6974-9}
\urldef\tempurl%
\url{https://doi.org/10.1145/3357236.3395485}
\showDOI{\tempurl}


\bibitem[Cheng et~al\mbox{.}(2023)]%
        {cheng_eviper_2023}
\bibfield{author}{\bibinfo{person}{Hsin Cheng}, \bibinfo{person}{Zhiwu Zheng},
  \bibinfo{person}{Prakhar Kumar}, \bibinfo{person}{Wali Afridi},
  \bibinfo{person}{Ben Kim}, \bibinfo{person}{Sigurd Wagner},
  \bibinfo{person}{Naveen Verma}, \bibinfo{person}{James~C. Sturm}, {and}
  \bibinfo{person}{Minjie Chen}.} \bibinfo{year}{2023}\natexlab{}.
\newblock \showarticletitle{{eViper}: {A} {Scalable} {Platform} for
  {Untethered} {Modular} {Soft} {Robots}}.
\newblock  (\bibinfo{year}{2023}).
\newblock
\urldef\tempurl%
\url{https://doi.org/10.48550/ARXIV.2303.01676}
\showDOI{\tempurl}
\newblock
\shownote{Publisher: [object Object] Version Number: 2}.


\bibitem[Deng et~al\mbox{.}(2014)]%
        {deng_tango_2014}
\bibfield{author}{\bibinfo{person}{Ying Deng}, \bibinfo{person}{Alissa~N.
  Antle}, {and} \bibinfo{person}{Carman Neustaedter}.}
  \bibinfo{year}{2014}\natexlab{}.
\newblock \showarticletitle{Tango cards: a card-based design tool for informing
  the design of tangible learning games}. In
  \bibinfo{booktitle}{\emph{Proceedings of the 2014 conference on {Designing}
  interactive systems}}. \bibinfo{publisher}{ACM}, \bibinfo{address}{Vancouver
  BC Canada}, \bibinfo{pages}{695--704}.
\newblock
\showISBNx{978-1-4503-2902-6}
\urldef\tempurl%
\url{https://doi.org/10.1145/2598510.2598601}
\showDOI{\tempurl}


\bibitem[Elsayed-Ali et~al\mbox{.}(2023)]%
        {elsayed-ali_responsible_2023}
\bibfield{author}{\bibinfo{person}{Salma Elsayed-Ali}, \bibinfo{person}{Sara~E
  Berger}, \bibinfo{person}{Vagner Figueredo~De Santana}, {and}
  \bibinfo{person}{Juana~Catalina Becerra~Sandoval}.}
  \bibinfo{year}{2023}\natexlab{}.
\newblock \showarticletitle{Responsible \& {Inclusive} {Cards}: {An} {Online}
  {Card} {Tool} to {Promote} {Critical} {Reflection} in {Technology} {Industry}
  {Work} {Practices}}. In \bibinfo{booktitle}{\emph{Proceedings of the 2023
  {CHI} {Conference} on {Human} {Factors} in {Computing} {Systems}}}.
  \bibinfo{publisher}{ACM}, \bibinfo{address}{Hamburg Germany},
  \bibinfo{pages}{1--14}.
\newblock
\showISBNx{978-1-4503-9421-5}
\urldef\tempurl%
\url{https://doi.org/10.1145/3544548.3580771}
\showDOI{\tempurl}


\bibitem[Fedosov et~al\mbox{.}(2019)]%
        {fedosov_sharing_2019}
\bibfield{author}{\bibinfo{person}{Anton Fedosov}, \bibinfo{person}{Masako
  Kitazaki}, \bibinfo{person}{William Odom}, {and} \bibinfo{person}{Marc
  Langheinrich}.} \bibinfo{year}{2019}\natexlab{}.
\newblock \showarticletitle{Sharing {Economy} {Design} {Cards}}. In
  \bibinfo{booktitle}{\emph{Proceedings of the 2019 {CHI} {Conference} on
  {Human} {Factors} in {Computing} {Systems}}}. \bibinfo{publisher}{ACM},
  \bibinfo{address}{Glasgow Scotland Uk}, \bibinfo{pages}{1--14}.
\newblock
\showISBNx{978-1-4503-5970-2}
\urldef\tempurl%
\url{https://doi.org/10.1145/3290605.3300375}
\showDOI{\tempurl}


\bibitem[Felton et~al\mbox{.}(2014)]%
        {felton2014method}
\bibfield{author}{\bibinfo{person}{Samuel Felton}, \bibinfo{person}{Michael
  Tolley}, \bibinfo{person}{Erik Demaine}, \bibinfo{person}{Daniela Rus}, {and}
  \bibinfo{person}{Robert Wood}.} \bibinfo{year}{2014}\natexlab{}.
\newblock \showarticletitle{A method for building self-folding machines}.
\newblock \bibinfo{journal}{\emph{Science}} \bibinfo{volume}{345},
  \bibinfo{number}{6197} (\bibinfo{year}{2014}), \bibinfo{pages}{644--646}.
\newblock


\bibitem[Firouzeh and Paik(2015)]%
        {firouzeh2015robogami}
\bibfield{author}{\bibinfo{person}{Amir Firouzeh} {and} \bibinfo{person}{Jamie
  Paik}.} \bibinfo{year}{2015}\natexlab{}.
\newblock \showarticletitle{Robogami: A fully integrated low-profile robotic
  origami}.
\newblock \bibinfo{journal}{\emph{Journal of Mechanisms and Robotics}}
  \bibinfo{volume}{7}, \bibinfo{number}{2} (\bibinfo{year}{2015}),
  \bibinfo{pages}{021009}.
\newblock


\bibitem[Friedman and Hendry(2012)]%
        {friedman_envisioning_2012}
\bibfield{author}{\bibinfo{person}{Batya Friedman} {and} \bibinfo{person}{David
  Hendry}.} \bibinfo{year}{2012}\natexlab{}.
\newblock \showarticletitle{The envisioning cards: a toolkit for catalyzing
  humanistic and technical imaginations}. In
  \bibinfo{booktitle}{\emph{Proceedings of the {SIGCHI} {Conference} on {Human}
  {Factors} in {Computing} {Systems}}}. \bibinfo{publisher}{ACM},
  \bibinfo{address}{Austin Texas USA}, \bibinfo{pages}{1145--1148}.
\newblock
\showISBNx{978-1-4503-1015-4}
\urldef\tempurl%
\url{https://doi.org/10.1145/2207676.2208562}
\showDOI{\tempurl}


\bibitem[Golick(1973)]%
        {golick_deal_1973}
\bibfield{author}{\bibinfo{person}{Margie Golick}.}
  \bibinfo{year}{1973}\natexlab{}.
\newblock \showarticletitle{Deal {Me} {In}! {The} {Use} of {Playing} {Cards} in
  {Learning} and {Teaching}.}
\newblock  (\bibinfo{year}{1973}).
\newblock
\newblock
\shownote{Publisher: ERIC}.


\bibitem[Gomes et~al\mbox{.}(2013)]%
        {gomes2013morephone}
\bibfield{author}{\bibinfo{person}{Antonio Gomes}, \bibinfo{person}{Andrea
  Nesbitt}, {and} \bibinfo{person}{Roel Vertegaal}.}
  \bibinfo{year}{2013}\natexlab{}.
\newblock \showarticletitle{MorePhone: a study of actuated shape deformations
  for flexible thin-film smartphone notifications}. In
  \bibinfo{booktitle}{\emph{Proceedings of the SIGCHI Conference on Human
  Factors in Computing Systems}}. \bibinfo{pages}{583--592}.
\newblock


\bibitem[Hao et~al\mbox{.}(2023)]%
        {hao_controlling_2023}
\bibfield{author}{\bibinfo{person}{Zhijian Hao}, \bibinfo{person}{Siddharth
  Mayya}, \bibinfo{person}{Gennaro Notomista}, \bibinfo{person}{Seth
  Hutchinson}, \bibinfo{person}{Magnus Egerstedt}, {and}
  \bibinfo{person}{Azadeh Ansari}.} \bibinfo{year}{2023}\natexlab{}.
\newblock \showarticletitle{Controlling {Collision}-{Induced} {Aggregations} in
  a {Swarm} of {Micro} {Bristle} {Robots}}.
\newblock \bibinfo{journal}{\emph{IEEE Transactions on Robotics}}
  \bibinfo{volume}{39}, \bibinfo{number}{1} (\bibinfo{date}{Feb.}
  \bibinfo{year}{2023}), \bibinfo{pages}{590--604}.
\newblock
\showISSN{1552-3098, 1941-0468}
\urldef\tempurl%
\url{https://doi.org/10.1109/TRO.2022.3189846}
\showDOI{\tempurl}


\bibitem[Heibeck et~al\mbox{.}(2015)]%
        {heibeck2015unimorph}
\bibfield{author}{\bibinfo{person}{Felix Heibeck}, \bibinfo{person}{Basheer
  Tome}, \bibinfo{person}{Clark Della~Silva}, {and} \bibinfo{person}{Hiroshi
  Ishii}.} \bibinfo{year}{2015}\natexlab{}.
\newblock \showarticletitle{uniMorph: Fabricating thin film composites for
  shape-changing interfaces}. In \bibinfo{booktitle}{\emph{Proceedings of the
  28th Annual ACM Symposium on User Interface Software \& Technology}}.
  \bibinfo{pages}{233--242}.
\newblock


\bibitem[Hoffmann(1973)]%
        {hoffmann1973playing}
\bibfield{author}{\bibinfo{person}{D. Hoffmann}.}
  \bibinfo{year}{1973}\natexlab{}.
\newblock \bibinfo{booktitle}{\emph{The Playing Card: An Illustrated History}}.
\newblock \bibinfo{publisher}{Edition Leipzig}.
\newblock
\showISBNx{9780904000030}
\showLCCN{74175507}
\urldef\tempurl%
\url{https://books.google.com/books?id=4_QYDQEACAAJ}
\showURL{%
\tempurl}


\bibitem[Holman et~al\mbox{.}(2005)]%
        {holman_paper_2005}
\bibfield{author}{\bibinfo{person}{David Holman}, \bibinfo{person}{Roel
  Vertegaal}, \bibinfo{person}{Mark Altosaar}, \bibinfo{person}{Nikolaus
  Troje}, {and} \bibinfo{person}{Derek Johns}.}
  \bibinfo{year}{2005}\natexlab{}.
\newblock \showarticletitle{Paper windows: interaction techniques for digital
  paper}. In \bibinfo{booktitle}{\emph{Proceedings of the {SIGCHI} conference
  on {Human} factors in computing systems}}. \bibinfo{pages}{591--599}.
\newblock


\bibitem[Hsieh et~al\mbox{.}(2023)]%
        {hsieh_what_2023}
\bibfield{author}{\bibinfo{person}{Gary Hsieh}, \bibinfo{person}{Brett~A.
  Halperin}, \bibinfo{person}{Evan Schmitz}, \bibinfo{person}{Yen~Nee Chew},
  {and} \bibinfo{person}{Yuan-Chi Tseng}.} \bibinfo{year}{2023}\natexlab{}.
\newblock \showarticletitle{What is in the {Cards}: {Exploring} {Uses},
  {Patterns}, and {Trends} in {Design} {Cards}}. In
  \bibinfo{booktitle}{\emph{Proceedings of the 2023 {CHI} {Conference} on
  {Human} {Factors} in {Computing} {Systems}}}. \bibinfo{publisher}{ACM},
  \bibinfo{address}{Hamburg Germany}, \bibinfo{pages}{1--18}.
\newblock
\showISBNx{978-1-4503-9421-5}
\urldef\tempurl%
\url{https://doi.org/10.1145/3544548.3580712}
\showDOI{\tempurl}


\bibitem[Ishii et~al\mbox{.}(2015)]%
        {ishii_transform_2015}
\bibfield{author}{\bibinfo{person}{Hiroshi Ishii}, \bibinfo{person}{Daniel
  Leithinger}, \bibinfo{person}{Sean Follmer}, \bibinfo{person}{Amit Zoran},
  \bibinfo{person}{Philipp Schoessler}, {and} \bibinfo{person}{Jared Counts}.}
  \bibinfo{year}{2015}\natexlab{}.
\newblock \showarticletitle{{TRANSFORM}: {Embodiment} of "{Radical} {Atoms}" at
  {Milano} {Design} {Week}}. In \bibinfo{booktitle}{\emph{Proceedings of the
  33rd {Annual} {ACM} {Conference} {Extended} {Abstracts} on {Human} {Factors}
  in {Computing} {Systems}}}. \bibinfo{publisher}{ACM}, \bibinfo{address}{Seoul
  Republic of Korea}, \bibinfo{pages}{687--694}.
\newblock
\showISBNx{978-1-4503-3146-3}
\urldef\tempurl%
\url{https://doi.org/10.1145/2702613.2702969}
\showDOI{\tempurl}


\bibitem[Jaquier et~al\mbox{.}(2024)]%
        {jaquier2024transferlearningroboticsupcoming}
\bibfield{author}{\bibinfo{person}{Noémie Jaquier},
  \bibinfo{person}{Michael~C. Welle}, \bibinfo{person}{Andrej Gams},
  \bibinfo{person}{Kunpeng Yao}, \bibinfo{person}{Bernardo Fichera},
  \bibinfo{person}{Aude Billard}, \bibinfo{person}{Aleš Ude},
  \bibinfo{person}{Tamim Asfour}, {and} \bibinfo{person}{Danica Kragic}.}
  \bibinfo{year}{2024}\natexlab{}.
\newblock \bibinfo{title}{Transfer Learning in Robotics: An Upcoming
  Breakthrough? A Review of Promises and Challenges}.
\newblock
\newblock
\showeprint[arxiv]{2311.18044}~[cs.RO]
\urldef\tempurl%
\url{https://arxiv.org/abs/2311.18044}
\showURL{%
\tempurl}


\bibitem[Kilic~Afsar et~al\mbox{.}(2021)]%
        {kilic_afsar_omnifiber_2021}
\bibfield{author}{\bibinfo{person}{Ozgun Kilic~Afsar}, \bibinfo{person}{Ali
  Shtarbanov}, \bibinfo{person}{Hila Mor}, \bibinfo{person}{Ken Nakagaki},
  \bibinfo{person}{Jack Forman}, \bibinfo{person}{Karen Modrei},
  \bibinfo{person}{Seung~Hee Jeong}, \bibinfo{person}{Klas Hjort},
  \bibinfo{person}{Kristina Höök}, {and} \bibinfo{person}{Hiroshi Ishii}.}
  \bibinfo{year}{2021}\natexlab{}.
\newblock \showarticletitle{{OmniFiber}: {Integrated} {Fluidic} {Fiber}
  {Actuators} for {Weaving} {Movement} based {Interactions} into the
  ‘{Fabric} of {Everyday} {Life}’}. In \bibinfo{booktitle}{\emph{The 34th
  {Annual} {ACM} {Symposium} on {User} {Interface} {Software} and
  {Technology}}}. \bibinfo{publisher}{ACM}, \bibinfo{address}{Virtual Event
  USA}, \bibinfo{pages}{1010--1026}.
\newblock
\showISBNx{978-1-4503-8635-7}
\urldef\tempurl%
\url{https://doi.org/10.1145/3472749.3474802}
\showDOI{\tempurl}


\bibitem[Kirshenbaum and Robertson(2018)]%
        {kirshenbaum_pepa_2018}
\bibfield{author}{\bibinfo{person}{Nurit Kirshenbaum} {and}
  \bibinfo{person}{Scott Robertson}.} \bibinfo{year}{2018}\natexlab{}.
\newblock \showarticletitle{{PEPA} {Deck}: {Bringing} {Interactivity} to
  {Playing} {Cards}}. In \bibinfo{booktitle}{\emph{Proceedings of the 2018
  {Annual} {Symposium} on {Computer}-{Human} {Interaction} in {Play}
  {Companion} {Extended} {Abstracts}}}. \bibinfo{publisher}{ACM},
  \bibinfo{address}{Melbourne VIC Australia}, \bibinfo{pages}{479--486}.
\newblock
\showISBNx{978-1-4503-5968-9}
\urldef\tempurl%
\url{https://doi.org/10.1145/3270316.3271521}
\showDOI{\tempurl}


\bibitem[Knapp(1996)]%
        {knapp_learning_1996}
\bibfield{author}{\bibinfo{person}{Thomas~R Knapp}.}
  \bibinfo{year}{1996}\natexlab{}.
\newblock \bibinfo{booktitle}{\emph{Learning statistics through playing
  cards}}.
\newblock \bibinfo{publisher}{Sage}.
\newblock


\bibitem[Koizumi et~al\mbox{.}(2010)]%
        {koizumi_animated_2010}
\bibfield{author}{\bibinfo{person}{Naoya Koizumi}, \bibinfo{person}{Kentaro
  Yasu}, \bibinfo{person}{Angela Liu}, \bibinfo{person}{Maki Sugimoto}, {and}
  \bibinfo{person}{Masahiko Inami}.} \bibinfo{year}{2010}\natexlab{}.
\newblock \showarticletitle{Animated paper: a moving prototyping platform}. In
  \bibinfo{booktitle}{\emph{Adjunct proceedings of the 23nd annual {ACM}
  symposium on {User} interface software and technology}}.
  \bibinfo{publisher}{ACM}, \bibinfo{address}{New York New York USA},
  \bibinfo{pages}{389--390}.
\newblock
\showISBNx{978-1-4503-0462-7}
\urldef\tempurl%
\url{https://doi.org/10.1145/1866218.1866234}
\showDOI{\tempurl}


\bibitem[Kolvenbag et~al\mbox{.}(2022)]%
        {kolvenbag_rapid_2022}
\bibfield{author}{\bibinfo{person}{Jay Kolvenbag}, \bibinfo{person}{Miguel
  Bruns}, {and} \bibinfo{person}{Amy Winters}.}
  \bibinfo{year}{2022}\natexlab{}.
\newblock \showarticletitle{Rapid {Prototyping} {Dynamic} {Robotic} {Fibers}
  for {Tunable} {Movement}}. In \bibinfo{booktitle}{\emph{Adjunct {Proceedings}
  of the 35th {Annual} {ACM} {Symposium} on {User} {Interface} {Software} and
  {Technology}}}. \bibinfo{publisher}{ACM}, \bibinfo{address}{Bend OR USA},
  \bibinfo{pages}{1--4}.
\newblock
\showISBNx{978-1-4503-9321-8}
\urldef\tempurl%
\url{https://doi.org/10.1145/3526114.3558696}
\showDOI{\tempurl}


\bibitem[Lam et~al\mbox{.}(2006)]%
        {albert2006art}
\bibfield{author}{\bibinfo{person}{Albert H.~T. Lam}, \bibinfo{person}{Kevin
  C.~H. Chow}, \bibinfo{person}{Edward H.~H. Yau}, {and}
  \bibinfo{person}{Michael~R. Lyu}.} \bibinfo{year}{2006}\natexlab{}.
\newblock \showarticletitle{ART: augmented reality table for interactive
  trading card game}. In \bibinfo{booktitle}{\emph{Proceedings of the 2006 ACM
  International Conference on Virtual Reality Continuum and Its Applications}}
  (Hong Kong, China) \emph{(\bibinfo{series}{VRCIA '06})}.
  \bibinfo{publisher}{Association for Computing Machinery},
  \bibinfo{address}{New York, NY, USA}, \bibinfo{pages}{357–360}.
\newblock
\showISBNx{1595933247}
\urldef\tempurl%
\url{https://doi.org/10.1145/1128923.1128987}
\showDOI{\tempurl}


\bibitem[Laput et~al\mbox{.}(2016)]%
        {laput_viband_2016}
\bibfield{author}{\bibinfo{person}{Gierad Laput}, \bibinfo{person}{Robert
  Xiao}, {and} \bibinfo{person}{Chris Harrison}.}
  \bibinfo{year}{2016}\natexlab{}.
\newblock \showarticletitle{{ViBand}: {High}-{Fidelity} {Bio}-{Acoustic}
  {Sensing} {Using} {Commodity} {Smartwatch} {Accelerometers}}. In
  \bibinfo{booktitle}{\emph{Proceedings of the 29th {Annual} {Symposium} on
  {User} {Interface} {Software} and {Technology}}}
  \emph{(\bibinfo{series}{{UIST} '16})}. \bibinfo{publisher}{Association for
  Computing Machinery}, \bibinfo{address}{New York, NY, USA},
  \bibinfo{pages}{321--333}.
\newblock
\showISBNx{978-1-4503-4189-9}
\urldef\tempurl%
\url{https://doi.org/10.1145/2984511.2984582}
\showDOI{\tempurl}
\newblock
\shownote{event-place: Tokyo, Japan}.


\bibitem[Le~Goc et~al\mbox{.}(2016)]%
        {le_goc_zooids_2016}
\bibfield{author}{\bibinfo{person}{Mathieu Le~Goc},
  \bibinfo{person}{Lawrence~H. Kim}, \bibinfo{person}{Ali Parsaei},
  \bibinfo{person}{Jean-Daniel Fekete}, \bibinfo{person}{Pierre Dragicevic},
  {and} \bibinfo{person}{Sean Follmer}.} \bibinfo{year}{2016}\natexlab{}.
\newblock \showarticletitle{Zooids: {Building} {Blocks} for {Swarm} {User}
  {Interfaces}}. In \bibinfo{booktitle}{\emph{Proceedings of the 29th {Annual}
  {Symposium} on {User} {Interface} {Software} and {Technology}}}.
  \bibinfo{publisher}{ACM}, \bibinfo{address}{Tokyo Japan},
  \bibinfo{pages}{97--109}.
\newblock
\showISBNx{978-1-4503-4189-9}
\urldef\tempurl%
\url{https://doi.org/10.1145/2984511.2984547}
\showDOI{\tempurl}


\bibitem[Lomas et~al\mbox{.}(2021)]%
        {lomas_design_2021}
\bibfield{author}{\bibinfo{person}{James~Derek Lomas}, \bibinfo{person}{Mihovil
  Karac}, {and} \bibinfo{person}{Mathieu Gielen}.}
  \bibinfo{year}{2021}\natexlab{}.
\newblock \showarticletitle{Design {Space} {Cards}: {Using} a {Card} {Deck} to
  {Navigate} the {Design} {Space} of {Interactive} {Play}}.
\newblock \bibinfo{journal}{\emph{Proceedings of the ACM on Human-Computer
  Interaction}} \bibinfo{volume}{5}, \bibinfo{number}{CHI PLAY}
  (\bibinfo{date}{Oct.} \bibinfo{year}{2021}), \bibinfo{pages}{1--21}.
\newblock
\showISSN{2573-0142}
\urldef\tempurl%
\url{https://doi.org/10.1145/3474654}
\showDOI{\tempurl}


\bibitem[Lucero and Arrasvuori(2010)]%
        {lucero_plex_2010}
\bibfield{author}{\bibinfo{person}{Andrés Lucero} {and} \bibinfo{person}{Juha
  Arrasvuori}.} \bibinfo{year}{2010}\natexlab{}.
\newblock \showarticletitle{\textit{{PLEX} {Cards}}: a source of inspiration
  when designing for playfulness}. In \bibinfo{booktitle}{\emph{Proceedings of
  the 3rd {International} {Conference} on {Fun} and {Games}}}.
  \bibinfo{publisher}{ACM}, \bibinfo{address}{Leuven Belgium},
  \bibinfo{pages}{28--37}.
\newblock
\showISBNx{978-1-60558-907-7}
\urldef\tempurl%
\url{https://doi.org/10.1145/1823818.1823821}
\showDOI{\tempurl}


\bibitem[Mayes and Markantonakis(2008)]%
        {mayes_smart_2008}
\bibfield{author}{\bibinfo{person}{Keith~E Mayes} {and}
  \bibinfo{person}{Konstantinos Markantonakis}.}
  \bibinfo{year}{2008}\natexlab{}.
\newblock \bibinfo{booktitle}{\emph{Smart cards, tokens, security and
  applications}}. Vol.~\bibinfo{volume}{1}.
\newblock \bibinfo{publisher}{Springer}.
\newblock


\bibitem[Nakagaki et~al\mbox{.}(2016)]%
        {nakagaki_chainform_2016}
\bibfield{author}{\bibinfo{person}{Ken Nakagaki}, \bibinfo{person}{Artem
  Dementyev}, \bibinfo{person}{Sean Follmer}, \bibinfo{person}{Joseph~A.
  Paradiso}, {and} \bibinfo{person}{Hiroshi Ishii}.}
  \bibinfo{year}{2016}\natexlab{}.
\newblock \showarticletitle{{ChainFORM}: {A} {Linear} {Integrated} {Modular}
  {Hardware} {System} for {Shape} {Changing} {Interfaces}}. In
  \bibinfo{booktitle}{\emph{Proceedings of the 29th {Annual} {Symposium} on
  {User} {Interface} {Software} and {Technology}}}. \bibinfo{publisher}{ACM},
  \bibinfo{address}{Tokyo Japan}, \bibinfo{pages}{87--96}.
\newblock
\showISBNx{978-1-4503-4189-9}
\urldef\tempurl%
\url{https://doi.org/10.1145/2984511.2984587}
\showDOI{\tempurl}


\bibitem[Nakagaki et~al\mbox{.}(2015)]%
        {nakagaki_lineform_2015}
\bibfield{author}{\bibinfo{person}{Ken Nakagaki}, \bibinfo{person}{Sean
  Follmer}, {and} \bibinfo{person}{Hiroshi Ishii}.}
  \bibinfo{year}{2015}\natexlab{}.
\newblock \showarticletitle{{LineFORM}: {Actuated} {Curve} {Interfaces} for
  {Display}, {Interaction}, and {Constraint}}. In
  \bibinfo{booktitle}{\emph{Proceedings of the 28th {Annual} {ACM} {Symposium}
  on {User} {Interface} {Software} \& {Technology}}}. \bibinfo{publisher}{ACM},
  \bibinfo{address}{Charlotte NC USA}, \bibinfo{pages}{333--339}.
\newblock
\showISBNx{978-1-4503-3779-3}
\urldef\tempurl%
\url{https://doi.org/10.1145/2807442.2807452}
\showDOI{\tempurl}


\bibitem[Nakagaki et~al\mbox{.}(2020)]%
        {nakagaki2020hermits}
\bibfield{author}{\bibinfo{person}{Ken Nakagaki}, \bibinfo{person}{Joanne
  Leong}, \bibinfo{person}{Jordan~L Tappa}, \bibinfo{person}{Jo{\~a}o Wilbert},
  {and} \bibinfo{person}{Hiroshi Ishii}.} \bibinfo{year}{2020}\natexlab{}.
\newblock \showarticletitle{Hermits: Dynamically reconfiguring the
  interactivity of self-propelled tuis with mechanical shell add-ons}. In
  \bibinfo{booktitle}{\emph{Proceedings of the 33rd Annual ACM Symposium on
  User Interface Software and Technology}}. \bibinfo{pages}{882--896}.
\newblock


\bibitem[Nakata(2019)]%
        {nakata2019learning}
\bibfield{author}{\bibinfo{person}{Tatsuya Nakata}.}
  \bibinfo{year}{2019}\natexlab{}.
\newblock \showarticletitle{Learning words with flash cards and word cards}.
\newblock \bibinfo{journal}{\emph{The Routledge handbook of vocabulary
  studies}} (\bibinfo{year}{2019}), \bibinfo{pages}{304--319}.
\newblock


\bibitem[Odenweller et~al\mbox{.}(1998)]%
        {odenweller1998educational}
\bibfield{author}{\bibinfo{person}{Cynthia~M Odenweller},
  \bibinfo{person}{Christopher~T Hsu}, {and} \bibinfo{person}{Stephen~E
  DiCarlo}.} \bibinfo{year}{1998}\natexlab{}.
\newblock \showarticletitle{Educational card games for understanding
  gastrointestinal physiology.}
\newblock \bibinfo{journal}{\emph{Advances in Physiology Education}}
  \bibinfo{volume}{275}, \bibinfo{number}{6} (\bibinfo{year}{1998}),
  \bibinfo{pages}{S78}.
\newblock


\bibitem[Ogata and Fukumoto(2015)]%
        {ogata_fluxpaper_2015}
\bibfield{author}{\bibinfo{person}{Masa Ogata} {and} \bibinfo{person}{Masaaki
  Fukumoto}.} \bibinfo{year}{2015}\natexlab{}.
\newblock \showarticletitle{{FluxPaper}: {Reinventing} {Paper} with {Dynamic}
  {Actuation} {Powered} by {Magnetic} {Flux}}. In
  \bibinfo{booktitle}{\emph{Proceedings of the 33rd {Annual} {ACM} {Conference}
  on {Human} {Factors} in {Computing} {Systems}}}. \bibinfo{publisher}{ACM},
  \bibinfo{address}{Seoul Republic of Korea}, \bibinfo{pages}{29--38}.
\newblock
\showISBNx{978-1-4503-3145-6}
\urldef\tempurl%
\url{https://doi.org/10.1145/2702123.2702516}
\showDOI{\tempurl}


\bibitem[Parshakova et~al\mbox{.}(2016)]%
        {parshakova_ratchair_2016}
\bibfield{author}{\bibinfo{person}{Tetiana Parshakova}, \bibinfo{person}{Minjoo
  Cho}, \bibinfo{person}{Alvaro Cassinelli}, {and} \bibinfo{person}{Daniel
  Saakes}.} \bibinfo{year}{2016}\natexlab{}.
\newblock \showarticletitle{Ratchair: furniture learns to move itself with
  vibration}. In \bibinfo{booktitle}{\emph{{ACM} {SIGGRAPH} 2016 {Emerging}
  {Technologies}}}. \bibinfo{publisher}{ACM}, \bibinfo{address}{Anaheim
  California}, \bibinfo{pages}{1--2}.
\newblock
\showISBNx{978-1-4503-4372-5}
\urldef\tempurl%
\url{https://doi.org/10.1145/2929464.2929473}
\showDOI{\tempurl}


\bibitem[Place(2005)]%
        {place_tarot_2005}
\bibfield{author}{\bibinfo{person}{Robert Place}.}
  \bibinfo{year}{2005}\natexlab{}.
\newblock \bibinfo{booktitle}{\emph{The {Tarot}: {History}, symbolism, and
  divination}}.
\newblock \bibinfo{publisher}{Penguin}.
\newblock


\bibitem[Rankl and Effing(2004)]%
        {rankl_smart_2004}
\bibfield{author}{\bibinfo{person}{Wolfgang Rankl} {and}
  \bibinfo{person}{Wolfgang Effing}.} \bibinfo{year}{2004}\natexlab{}.
\newblock \bibinfo{booktitle}{\emph{Smart card handbook}}.
\newblock \bibinfo{publisher}{John Wiley \& Sons}.
\newblock


\bibitem[Rogerson et~al\mbox{.}(2022)]%
        {rogerson_smeft_2022}
\bibfield{author}{\bibinfo{person}{Melissa~J. Rogerson},
  \bibinfo{person}{Lucy~A. Sparrow}, {and} \bibinfo{person}{Sophie~O.
  Freeman}.} \bibinfo{year}{2022}\natexlab{}.
\newblock \showarticletitle{The {SMeFT} {Decks}: {A} {Card}-{Based} {Ideation}
  {Tool} for {Designing} {Hybrid} {Digital} {Boardgames} for {Distanced}
  {Play}}. In \bibinfo{booktitle}{\emph{Proceedings of the 34th {Australian}
  {Conference} on {Human}-{Computer} {Interaction}}}. \bibinfo{publisher}{ACM},
  \bibinfo{address}{Canberra ACT Australia}, \bibinfo{pages}{298--309}.
\newblock
\showISBNx{9798400700248}
\urldef\tempurl%
\url{https://doi.org/10.1145/3572921.3572933}
\showDOI{\tempurl}


\bibitem[Roudaut et~al\mbox{.}(2013)]%
        {roudaut2013morphees}
\bibfield{author}{\bibinfo{person}{Anne Roudaut}, \bibinfo{person}{Abhijit
  Karnik}, \bibinfo{person}{Markus L{\"o}chtefeld}, {and}
  \bibinfo{person}{Sriram Subramanian}.} \bibinfo{year}{2013}\natexlab{}.
\newblock \showarticletitle{Morphees: toward high" shape resolution" in
  self-actuated flexible mobile devices}. In
  \bibinfo{booktitle}{\emph{Proceedings of the SIGCHI conference on human
  factors in computing systems}}. \bibinfo{pages}{593--602}.
\newblock


\bibitem[Roy and Warren(2019)]%
        {roy_card-based_2019}
\bibfield{author}{\bibinfo{person}{Robin Roy} {and} \bibinfo{person}{James~P
  Warren}.} \bibinfo{year}{2019}\natexlab{}.
\newblock \showarticletitle{Card-based design tools: {A} review and analysis of
  155 card decks for designers and designing}.
\newblock \bibinfo{journal}{\emph{Design Studies}}  \bibinfo{volume}{63}
  (\bibinfo{year}{2019}), \bibinfo{pages}{125--154}.
\newblock
\newblock
\shownote{Publisher: Elsevier}.


\bibitem[Rubenstein et~al\mbox{.}(2012)]%
        {rubenstein_kilobot_2012}
\bibfield{author}{\bibinfo{person}{Michael Rubenstein},
  \bibinfo{person}{Christian Ahler}, {and} \bibinfo{person}{Radhika Nagpal}.}
  \bibinfo{year}{2012}\natexlab{}.
\newblock \showarticletitle{Kilobot: {A} low cost scalable robot system for
  collective behaviors}. In \bibinfo{booktitle}{\emph{2012 {IEEE}
  {International} {Conference} on {Robotics} and {Automation}}}.
  \bibinfo{publisher}{IEEE}, \bibinfo{address}{St Paul, MN, USA},
  \bibinfo{pages}{3293--3298}.
\newblock
\showISBNx{978-1-4673-1405-3 978-1-4673-1403-9 978-1-4673-1578-4
  978-1-4673-1404-6}
\urldef\tempurl%
\url{https://doi.org/10.1109/ICRA.2012.6224638}
\showDOI{\tempurl}


\bibitem[Römer and Domnitcheva(2002)]%
        {romer_smart_2002}
\bibfield{author}{\bibinfo{person}{Kay Römer} {and} \bibinfo{person}{Svetlana
  Domnitcheva}.} \bibinfo{year}{2002}\natexlab{}.
\newblock \showarticletitle{Smart {Playing} {Cards}: {A} {Ubiquitous}
  {Computing} {Game}}.
\newblock \bibinfo{journal}{\emph{Personal and Ubiquitous Computing}}
  \bibinfo{volume}{6}, \bibinfo{number}{5-6} (\bibinfo{date}{Dec.}
  \bibinfo{year}{2002}), \bibinfo{pages}{371--377}.
\newblock
\showISSN{16174909}
\urldef\tempurl%
\url{https://doi.org/10.1007/s007790200042}
\showDOI{\tempurl}


\bibitem[Scalise et~al\mbox{.}(2022)]%
        {scalise_deal_2022}
\bibfield{author}{\bibinfo{person}{Nicole~R. Scalise}, \bibinfo{person}{Mary
  DePascale}, \bibinfo{person}{Nadia Tavassolie}, \bibinfo{person}{Claire
  McCown}, {and} \bibinfo{person}{Geetha~B. Ramani}.}
  \bibinfo{year}{2022}\natexlab{}.
\newblock \showarticletitle{Deal {Me} in: {Playing} {Cards} in the {Home} to
  {Learn} {Math}}.
\newblock \bibinfo{journal}{\emph{Education Sciences}} \bibinfo{volume}{12},
  \bibinfo{number}{3} (\bibinfo{date}{March} \bibinfo{year}{2022}),
  \bibinfo{pages}{190}.
\newblock
\showISSN{2227-7102}
\urldef\tempurl%
\url{https://doi.org/10.3390/educsci12030190}
\showDOI{\tempurl}


\bibitem[Schnorr(1991)]%
        {schnorr_efficient_1991}
\bibfield{author}{\bibinfo{person}{Claus-Peter Schnorr}.}
  \bibinfo{year}{1991}\natexlab{}.
\newblock \showarticletitle{Efficient signature generation by smart cards}.
\newblock \bibinfo{journal}{\emph{Journal of cryptology}}  \bibinfo{volume}{4}
  (\bibinfo{year}{1991}), \bibinfo{pages}{161--174}.
\newblock
\newblock
\shownote{Publisher: Springer}.


\bibitem[Shen et~al\mbox{.}(2021)]%
        {shen2021value}
\bibfield{author}{\bibinfo{person}{Hong Shen}, \bibinfo{person}{Wesley~H Deng},
  \bibinfo{person}{Aditi Chattopadhyay}, \bibinfo{person}{Zhiwei~Steven Wu},
  \bibinfo{person}{Xu Wang}, {and} \bibinfo{person}{Haiyi Zhu}.}
  \bibinfo{year}{2021}\natexlab{}.
\newblock \showarticletitle{Value cards: An educational toolkit for teaching
  social impacts of machine learning through deliberation}. In
  \bibinfo{booktitle}{\emph{Proceedings of the 2021 ACM conference on fairness,
  accountability, and transparency}}. \bibinfo{pages}{850--861}.
\newblock


\bibitem[Shepherd et~al\mbox{.}(2011)]%
        {shepherd2011multigait}
\bibfield{author}{\bibinfo{person}{Robert~F Shepherd}, \bibinfo{person}{Filip
  Ilievski}, \bibinfo{person}{Wonjae Choi}, \bibinfo{person}{Stephen~A Morin},
  \bibinfo{person}{Adam~A Stokes}, \bibinfo{person}{Aaron~D Mazzeo},
  \bibinfo{person}{Xin Chen}, \bibinfo{person}{Michael Wang}, {and}
  \bibinfo{person}{George~M Whitesides}.} \bibinfo{year}{2011}\natexlab{}.
\newblock \showarticletitle{Multigait soft robot}.
\newblock \bibinfo{journal}{\emph{Proceedings of the national academy of
  sciences}} \bibinfo{volume}{108}, \bibinfo{number}{51}
  (\bibinfo{year}{2011}), \bibinfo{pages}{20400--20403}.
\newblock


\bibitem[Smith and Munro(2009)]%
        {smith2009educational}
\bibfield{author}{\bibinfo{person}{DR Smith} {and} \bibinfo{person}{E Munro}.}
  \bibinfo{year}{2009}\natexlab{}.
\newblock \showarticletitle{Educational card games}.
\newblock \bibinfo{journal}{\emph{Physics Education}} \bibinfo{volume}{44},
  \bibinfo{number}{5} (\bibinfo{year}{2009}), \bibinfo{pages}{479}.
\newblock


\bibitem[Sony({[n.\,d.]})]%
        {toio}
\bibfield{author}{\bibinfo{person}{Sony}.} \bibinfo{year}{[n.\,d.]}\natexlab{}.
\newblock \showarticletitle{Toy Platform toio™}.
\newblock  (\bibinfo{year}{[n.\,d.]}).
\newblock
\urldef\tempurl%
\url{https://www.sony.com/en/SonyInfo/design/stories/toio/}
\showURL{%
\tempurl}


\bibitem[Tan et~al\mbox{.}(2018)]%
        {tan2018Sim2Real}
\bibfield{author}{\bibinfo{person}{Jie Tan}, \bibinfo{person}{Tingnan Zhang},
  \bibinfo{person}{Erwin Coumans}, \bibinfo{person}{Atil Iscen},
  \bibinfo{person}{Yunfei Bai}, \bibinfo{person}{Danijar Hafner},
  \bibinfo{person}{Steven Bohez}, {and} \bibinfo{person}{Vincent Vanhoucke}.}
  \bibinfo{year}{2018}\natexlab{}.
\newblock \showarticletitle{Sim-to-Real: Learning Agile Locomotion For
  Quadruped Robots}.
\newblock \bibinfo{journal}{\emph{CoRR}}  \bibinfo{volume}{abs/1804.10332}
  (\bibinfo{year}{2018}).
\newblock
\showeprint[arXiv]{1804.10332}
\urldef\tempurl%
\url{http://arxiv.org/abs/1804.10332}
\showURL{%
\tempurl}


\bibitem[Villar et~al\mbox{.}(2018)]%
        {villar2018project}
\bibfield{author}{\bibinfo{person}{Nicolas Villar}, \bibinfo{person}{Daniel
  Cletheroe}, \bibinfo{person}{Greg Saul}, \bibinfo{person}{Christian Holz},
  \bibinfo{person}{Tim Regan}, \bibinfo{person}{Oscar Salandin},
  \bibinfo{person}{Misha Sra}, \bibinfo{person}{Hui-Shyong Yeo},
  \bibinfo{person}{William Field}, {and} \bibinfo{person}{Haiyan Zhang}.}
  \bibinfo{year}{2018}\natexlab{}.
\newblock \showarticletitle{Project zanzibar: A portable and flexible tangible
  interaction platform}. In \bibinfo{booktitle}{\emph{Proceedings of the 2018
  CHI Conference on Human Factors in Computing Systems}}.
  \bibinfo{pages}{1--13}.
\newblock


\bibitem[Wilkinson(1895)]%
        {wilkinson_chinese_1895}
\bibfield{author}{\bibinfo{person}{William~H Wilkinson}.}
  \bibinfo{year}{1895}\natexlab{}.
\newblock \showarticletitle{Chinese origin of playing cards}.
\newblock \bibinfo{journal}{\emph{American Anthropologist}}
  \bibinfo{volume}{8}, \bibinfo{number}{1} (\bibinfo{year}{1895}),
  \bibinfo{pages}{61--78}.
\newblock
\newblock
\shownote{Publisher: JSTOR}.


\bibitem[Winfield et~al\mbox{.}(2007)]%
        {Winfield2007tpad}
\bibfield{author}{\bibinfo{person}{Laura Winfield}, \bibinfo{person}{John
  Glassmire}, \bibinfo{person}{J.~Edward Colgate}, {and}
  \bibinfo{person}{Michael Peshkin}.} \bibinfo{year}{2007}\natexlab{}.
\newblock \showarticletitle{T-PaD: Tactile Pattern Display through Variable
  Friction Reduction}. In \bibinfo{booktitle}{\emph{Second Joint EuroHaptics
  Conference and Symposium on Haptic Interfaces for Virtual Environment and
  Teleoperator Systems (WHC'07)}}. \bibinfo{pages}{421--426}.
\newblock
\urldef\tempurl%
\url{https://doi.org/10.1109/WHC.2007.105}
\showDOI{\tempurl}


\bibitem[Wit et~al\mbox{.}(2021)]%
        {toioblackjack}
\bibfield{author}{\bibinfo{person}{Justin Wit}, \bibinfo{person}{Dylan Schils},
  {and} \bibinfo{person}{Matthew Eicholtz}.} \bibinfo{year}{2021}\natexlab{}.
\newblock \showarticletitle{Tangiers Toios: Robotic Blackjack Dealers}.
\newblock  (\bibinfo{year}{2021}).
\newblock
\urldef\tempurl%
\url{https://www.youtube.com/watch?v=85TVs3F-DmM}
\showURL{%
\tempurl}


\bibitem[Wu et~al\mbox{.}(2009)]%
        {wu2009gesture}
\bibfield{author}{\bibinfo{person}{Jiahui Wu}, \bibinfo{person}{Gang Pan},
  \bibinfo{person}{Daqing Zhang}, \bibinfo{person}{Guande Qi}, {and}
  \bibinfo{person}{Shijian Li}.} \bibinfo{year}{2009}\natexlab{}.
\newblock \showarticletitle{Gesture recognition with a 3-d accelerometer}. In
  \bibinfo{booktitle}{\emph{Ubiquitous Intelligence and Computing: 6th
  International Conference, UIC 2009, Brisbane, Australia, July 7-9, 2009.
  Proceedings 6}}. Springer, \bibinfo{pages}{25--38}.
\newblock


\end{thebibliography}

\appendix

\end{document}